\documentclass[apj]{emulateapj}
\usepackage{graphicx,amsmath}
\usepackage{amssymb}
\usepackage{color}
\usepackage{natbib} 
\shorttitle{SFR, dust \& EBL}
\shortauthors{Khaire \& Srianand}

\begin{document}

\title{Star formation history, dust attenuation and extragalactic background light}

\author{Vikram Khaire and Raghunathan Srianand}
\affil{Inter-University Centre for Astronomy and Astrophysics (IUCAA), Post Bag 4, Pune 411007, India}
\email{Email: vikramk@iucaa.ernet.in}

\begin{abstract}
At any given epoch, the Extragalactic Background Light (EBL) carries imprints of 
integrated star formation activities in the universe till that epoch. On the other 
hand, in order to estimate the EBL, when direct observations are not possible, 
one requires an accurate estimation of the star formation rate density (SFRD) and 
the dust attenuation ($A_\nu$) in galaxies. Here, we present a `progressive fitting 
method' that determines the global average SFRD($z$) and $A_\nu$($z$) for any given 
extinction curve by using the available multi-wavelength multi-epoch galaxy luminosity 
function measurements. Using the available observations, we determine the best fitted 
combinations of SFRD($z$) and $A_\nu$($z$), in a simple fitting form, up to $z\sim8$ 
for five well known extinction curves. We find, irrespective of the extinction curve 
used, the $z$ at which the SFRD($z$) peaks is higher than the $z$ above which 
$A_\nu$($z$) begins to decline. For each case, we compute the EBL from ultra-violet 
to the far-infrared regime and the optical depth ($\tau_\gamma$) encountered by the 
high energy $\gamma$-rays due to pair production upon collisions with these EBL photons. 
We compare these with measurements of the local EBL, $\gamma$-ray horizon and 
$\tau_\gamma$ measurements using Fermi-LAT. All these and the comparison of independent 
SFRD($z$) and $A_\nu$($z$) measurements from the literature with our predictions favor 
the extinction curve similar to that of Large Magellanic cloud Supershell. 
\end{abstract}


\section{Introduction}
The extragalactic background light (EBL) at any epoch is a diffuse isotropic background 
radiation, defined here over the wavelength range 0.1 to 1000$\mu$m excluding the cosmic 
microwave background radiation (CMBR), believed to be contributed mainly by the sources 
such as galaxies and QSOs. The knowledge of how the intensity and shape of the EBL evolves 
is very important for understanding the galaxy evolution in the universe. 

Direct measurements of the EBL are possible only in the local universe 
\citep{Dwek98, Dole06,Matsuoka11}. However, there are large uncertainties associated with 
the removal of foreground contributions from the unresolved point sources and zodiacal light 
\citep[see][]{Hauser01}. The local EBL can be inferred by adding the light from resolved 
sources \citep{Madau00, Xu05, Hopwood10} but the convergence of the number of sources is 
in dispute \citep{Bernstein02,Levenson08}. However, with the aid of rapidly developing 
$\gamma$-ray astronomy, in principle, it is possible to place strong constraints on the 
intermediate redshift (z$<$2) EBL. 
  
The high energy $\gamma$-rays by interacting with the EBL photons can annihilate themselves 
and produce electron positron pairs. The byproduct of this interaction, ultra-relativistic 
electron positron pairs, are expected to inverse Compton scatter the CMBR and produce 
secondary $\gamma$-rays. These secondary $\gamma$-rays are not yet detected by the Fermi 
satellite implying either the presence of a small intergalactic magnetic field that scatters 
the produced pairs \citep[][]{Neronov10, Tavecchio11, Takahashi12, Arlen12} or the  
electromagnetic pair cascade dissipating pair beam energy in to the intergalactic medium  
\citep[IGM; see for e.g.,][]{Schlickeiser13, Miniati13}. Nevertheless, this process of 
pair production attenuates the $\gamma$-rays originating from distant sources while 
traveling through the IGM \citep{Gould66, Jelley66}. The amount of attenuation suffered by 
the $\gamma$-rays emitted by sources at different emission redshifts depends on the number 
density of the EBL photons encountered while traveling from the source to the earth. 
Thus a well measured $\gamma$-ray attenuation ($\tau_{\gamma}$) can be used to put 
constraints on the evolution of the shape and amplitude of the EBL. 
This was initially suggested by \citet{Stecker92} and the first few limits on the IR part of 
the EBL were placed by \citet{Dwek94} and \citet{deJager94} using TeV $\gamma$-ray 
observations of the blazar Mrk 421. 
 
With the aid of new generation ground based $\gamma$-ray Imagining Atmospheric Cherenkov 
telescopes (IACT) and the Fermi satellite, many high energy $\gamma$-ray sources have been 
detected. The observed spectrum of distant $\gamma$-ray sources are used to determine the 
$\tau_{\gamma}$. However, the difficulty in doing so arises from the fact that the 
intrinsic spectral energy distribution (SED) of each source is unknown. 
Recently, \citet{GammaSci} have circumvent this difficulty and reported the measurements of 
$\tau_{\gamma}$ up to $z\sim 1.5$ with the observed $\gamma$-ray energies from 10 to 500 
GeV using the stacked spectra of $\gamma$-ray blazars selected from the sample of objects 
observed with Large Area Telescope (LAT) on board the Fermi satellite. It is expected to 
find the $\tau_{\gamma}$ increasing with increasing redshift since these $\gamma$-rays 
travel longer distances through EBL photons to reach earth. The observation of this 
cosmological evolution in $\tau_{\gamma}$ has been reported recently by \citet{Sanchez13}. 
The $\gamma$-ray horizon for $\gamma$-ray photons with the observed energy $E_{\gamma}$ 
is defined as the emission redshift of $\gamma$-rays beyond which they encounter 
$\tau_{\gamma}>1$. Recently, by using a physically motivated modeling of intrinsic SEDs 
of 15 blazars, \citet{Dominguez13} reported the $\gamma$-ray horizon measurements. 
\citet{Dominguez13a} using such  $\gamma$-ray horizon measurements have demonstrated 
the capability of $\gamma$-ray astronomy to measure the Hubble constant. Recently, 
\citet{Scully14} and \citet{Stecker12} used their EBL model to constrain the redshift of 
$\gamma$-ray blazars. In some sense, the backbone of this rapidly developing $\gamma$-ray 
astronomy is the EBL. Therefore, it is very important to have an EBL estimates consistent 
with different observations over a large redshift range.

To estimate the EBL one needs the specific emissivity (or some times referred as luminosity 
density) at each frequency and redshift. The EBL in optical wavelengths is 
predominantly contributed by the stellar emission and in the infra-red (IR) 
wavelengths by dust emission from galaxies. Therefore, 
for the correct estimation of the EBL one has to determine the comoving specific galaxy 
emissivity at different frequencies and redshifts, $\rho_{\nu}(z)$, as accurately as 
possible. The EBL models are generally classified in to different categories depending on 
the method adopted to get the $\rho_{\nu}(z)$. For example, some of the models, say the 
first kind of models, start with simulating the galaxy evolution in the framework of 
standard cosmological model taking into account the dark matter halo formation and some 
prescription to relate baryons to star formation in them. These models then predict 
the $\rho_{\nu}(z)$ forward in time \citep{Primack05, Gilmore09, Gilmore12, Inoue13}. There 
are second type of models which construct the grid of $\rho_{\nu}(z)$ measurements in 
different wavebands and redshift and then apply the interpolation and the extrapolation to 
get the $\rho_{\nu}(z)$ at each $\nu$ and $z$ 
\citep{Stecker06, Franceschini08, Dominguez11, Stecker12, Helgason12}. There are third kind 
of models where the cosmic star formation history and the SED of stellar population of 
galaxies are convolved to get the $\rho_{\nu}(z)$ \citep{Kneiske04, Finke10, HM2012}. 
The $\rho_{\nu}(z)$ obtained in this way depends on star formation history of galaxies over 
the cosmic time and absorption and scattering by the dust present in them. Main uncertainties 
in the first and third approach are related to the the amount of dust corrections which is 
usually quantified by the dust attenuation magnitude $A_{\nu_0}$ at frequency $\nu_0$ and a 
wavelength dependent dust extinction curve. It is a general practice to assume a form of 
$A_{\nu_0}(z)$ and an extinction curve to get the $\rho_{\nu}(z)$ from the star formation 
history. Irrespective of the approach one adopts, all the methods are expected to reproduce 
the measured $\rho_{\nu}(z)$ using observed luminosity functions.

Here, in this paper, we address the issue of self consistently determining the dust 
correction and the star formation history which will reproduce the observed emissivity. 
We present a novel `progressive fitting method' which by using the $\rho_{\nu}(z)$, 
for a given extinction curve, determines a unique combination of cosmic star formation 
rate density (SFRD) and $A_{\nu_0}(z)$. We apply this method on observationally determined 
$\rho_{\nu}(z)$ using the available multi-wavelength multi-epoch galaxy data from the 
literature. We determine the combinations of SFRD($z$) and $A_{\nu_0}(z)$ for a set of 
five well known extinction curves and compare the results with different independent 
measurements of SFRD($z$) and $A_{\nu_0}(z)$ available in the literature. This allows us to 
determine the average extinction curve that can be used to convert the emissivity into 
the SFRD($z$). We provide the simple fitting forms of these combinations of SFRD($z$) 
and $A_{\nu_0}(z)$ for each extinction curve with their $1\sigma$ upper and lower limits. 
We self-consistently determine the amount of stellar light absorbed by dust with the help 
of these combinations of $A_{\nu_0}(z)$ and SFRD($z$) obtained for different extinction 
curves and then estimate the far infra-red (FIR) emission from  galaxies using the local 
galaxy FIR templates and the energy conservation arguments. In this way we obtain the 
specific emissivity from UV to FIR and then use standard prescription to calculate the EBL, 
the $\tau_{\gamma}$ and  the $\gamma$-ray horizon and compare these results with 
different available measurements. We conclude that the combination of $A_{\nu_0}(z)$ and 
SFRD($z$) obtained using the extinction curve of Large Magellanic cloud Supershell (LMC2) 
and the inferred local FIR emissivity are consistent with the different measurements and 
we call the EBL obtained using it as our fiducial model for the EBL. The EBL obtained in 
this way, by exploring different well known extinction curves and corresponding combinations 
of self-consistent $A_{\nu_0}(z)$ and SFRD($z$), includes better treatment of dust correction 
and gives a general picture of how the FIR part of the EBL depends on it. 

The outline of this paper is as follows. In section~\ref{sec.ebl} we present the standard 
radiative transfer equation used for calculating the EBL from the inferred emissivities. 
In section~\ref{sec.qso}, we describe the QSO contribution to the total emissivity used by us.
In section~\ref{sec.egal}, we describe the standard procedure to get the galaxy emissivity
using the SFRD($z$) and $A_{\nu_0}(z)$. In section~\ref{sec.method}, we summarize the galaxy 
emissivity measurements from the literature that we use in our study and describe our 
`progressive fitting' technique which determines a unique combination of $A_{\nu_0}(z)$ 
and SFRD($z$) for an assumed extinction curve. In section~\ref{sec.sfr}, we make a detailed 
comparison of SFRD($z$) and $A_{\nu_0}(z)$ obtained using our technique for five different 
extinction curves with those determined from the independent observations. We explain in 
detail the method used by us to calculate the FIR emissivity from galaxies in 
section~\ref{sec.fir}. Then we use these inferred galaxy emissivities to calculate the EBL 
at different $z$. We present our EBL predictions and compare them with the other EBL 
estimates from the literature in section~\ref{sec.eblcal}. In section~\ref{sec.gamma-tau}, 
we describe the basics of the pair production mechanism used for calculating the 
$\tau_{\gamma}$ for our EBL models and compare our results with the other independent 
measurements. We conclude with the discussion related to the uncertainties in estimating 
the star formation history, $A_{\nu_0}(z)$ and the EBL in section~\ref{sec.res} and summarize 
the results in section~\ref{sec.sum}. Throughout the paper we use cosmology with
$\Omega_{\lambda}$=0.7, $\Omega_{m}$=0.3 and $H_{0}$=70 km s$^{-1}$ Mpc$^{-1}$. 
%
\section{Cosmological radiative transfer}\label{sec.ebl}
In this section, we provide a general outline for the basic EBL calculations.
The number density of background photons at a frequency $ \nu_{0}$ and redshift $z_{0}$ 
is given by, 
%
\begin{equation}\label{Eq.num}
n(\nu_{0},z_{0})= \frac{4\pi J_{\nu_{0}}(z_{0})}{hc} \,\, ,
\end{equation}
%
where, $h$ is the Planck's constant and $J_{\nu_{0}}$ is the specific intensity of the EBL 
(in units of erg cm$^{\text{-2}}$ s$^{\text{-1}}$ Hz$^{\text{-1}}$ sr$^{\text{-1}}$) at a 
frequency $\nu_{0}$. Following the standard procedure \citep[see for example,][referred as 
HM12 from now onwards]{HM2012}, we assume that the QSOs and galaxies are the sole 
contributers to the EBL at all wavelengths. We do not consider contributions to the EBL 
from the non-standard sources like decaying dark matter or dark energy. From the observed 
luminosity functions of QSOs and galaxies at a redshift $z$ and a frequency $\nu$ 
one can calculate the proper space averaged specific volume emissivity $\epsilon_{\nu}(z)$ 
(in units of erg  s$^{\text{-1}}$ Hz$^{\text{-1}}$ Mpc$^{\text{-3}}$). 
Then, the radiative transfer equation, which gives the specific intensity 
$J_{\nu_{0}}(z_0)$ of the EBL as seen by an observer at a redshift $z_{0}$ and a frequency 
$\nu_{0}$, can be written as \citep{Peebles, HM96},
%
\begin{equation}\label{Eq.rad_t}
J_{\nu_{0}}(z_{0})=\frac{1}{4\pi}\int_{z_{0}}^{\infty}dz\,\frac{dl}{dz}\,\frac{(1+z_{0})^{3}}{(1+z)^{3}}\,\epsilon_{\nu}(z)\,e^{-\tau_{eff}(\nu_{0},z_{0},z)}.
\end{equation}
%
Here, $\frac{dl}{dz}$ is the cosmological Freidmann-Robertson-Walker (FRW) line element,
the $\nu=\nu_{0}(1+z)/(1+z_{0})$ is a frequency of the radiation originated from a  
redshift $z$ and $\tau_{eff}(\nu_{0},z_{0},z)$ is the effective IGM optical depth 
encountered by the radiation emitted at a frequency $\nu$ while traveling through the IGM 
from an emission redshift $z$ to a redshift $z_{0}$ where it has been observed at a 
frequency $\nu_{0}$. The hydrogen and helium gas present in the IGM and in galaxies 
dominate $\tau_{eff}$ at $\lambda \leq 0.091\mu$m through the photo-absorption. However, 
in the optical wavelengths it was believed that the main contribution to the opacity comes 
from the attenuation by the dust associated with high H~{\sc i} column density intervening 
systems. Based on the available QSO spectroscopic observations one can conclude that this 
effect is indeed negligible \citep[see,][]{Srianand97, York06, Frank10, Khare12, Menard12}.
Here, as we are interested in calculating the EBL at $\lambda > $  0.1$\mu m$, 
we will consider $\tau_{eff}$=0 in Eq.~\ref{Eq.rad_t}. This assumption has negligible effect 
on the computed EBL and it does not affect the $\tau_\gamma$ significantly over the 
$\gamma$-ray energy range of our interest.
%
%
\subsection{QSO contribution to emissivity}\label{sec.qso}
The proper specific volume emissivity of the radiating sources can be written as, 
\begin{equation}
\epsilon_{\nu}(z)=\epsilon_{\nu , \rm Q}(z)+ \epsilon_{\nu , \rm G}(z) \,\,, 
\end{equation}
%
where, $\epsilon_{\nu , \rm Q}(z)$ and $\epsilon_{\nu , \rm G}(z)$ are the proper specific 
volume emissivity of QSOs and galaxies, respectively. 
For QSOs, we use the parametric form for $\epsilon_{\nu , \rm Q}(z)$ as given in HM12 at 
$912\text{\AA}$ which is consistent with the QSO luminosity function of \citet{Hopkins}, 
%
\begin{equation}\label{Eqso}
\frac{\epsilon_{912, \rm Q}(z)}{(1+z)^{3}}=10^{24.6}\,(1+z)^{4.68}\,\frac{exp(-0.28z)}{exp(1.77z) + 26.3}\,\,\,,
\end{equation}
%
in units of ergs s$^{\text{-1}}$ Mpc$^{\text{-3}}$  Hz$^{\text{-1}}$. To get 
$\epsilon_{\nu , \rm Q}$ at different wavelengths, we use a SED given by the broken power law, 
$L_\nu\propto \nu^{-0.44}$ for $\lambda>$1300\AA~and $L_\nu\propto \nu^{-1.57}$ for 
$\lambda<$1300\AA~\citep{vanden,tefler}. It is well known that the stellar emission 
from  galaxies dominate the EBL in the optical regime in all redshifts. Therefore, we place 
more emphasis on the  estimating $\epsilon_{\nu , \rm G}(z)$ accurately. We discuss this in 
detail in the following section.
%
%
\subsection{Galaxy contribution to emissivity}\label{sec.egal}
\begin{table*}
\caption{Details of the observed galaxy luminosity functions used to get the $\rho_{\nu}$ in our study.}
\begin{center}
\begin{tabular}{l c c c c c c }
\hline              
\hline
 Reference & Waveband$^*$ & Redshift range &  Plotting Symbol$^\dagger$   \\
\hline
\citet{Schiminovich05}   & FUV        & 0.2-2.95     &      magenta triangle      \\
\citet{Reddy09}          & FUV        & 1.9-3.4      &      orange triangle       \\
\citet{Bouwens07}        & FUV        & 3.8-5.9      &      Red diamond            \\
\citet{Bouwens11}        & FUV        & 6.8-8        &      Red diamond            \\
\citet{Dahlen07}         & FUV        & 0.92-2.37    &      blue square           \\
                         & NUV        & 0.29-2.37    &      blue square           \\
\citet{Cucciati12}       & FUV        & 0.05-4.5     &      green circle          \\
                         & NUV        & 0.05-3.5     &      green circle          \\
\citet{Tresse07} & FUV, NUV, U, V, B, R, I & 0.05-2  &      red circle            \\
\citet{Wyder05}          & NUV        & 0.055        &      black triangle        \\
\citet{Faber07}          & B          & 0.2-1.2      &      orange star           \\
\citet{Dahlen05}         & U, B, R    & 0.1-2        &      blue square           \\
                         & J          & 0.1-1        &      blue square           \\
\citet{Stefanon13}       & J, H       & 1.5-3.5      &      green square          \\
\citet{Pozzetti03}       & J, K       & 0.2-1.3      &      orange triangle       \\
\citet{Arnouts07}        & K          & 0.2-2        &      green diamond          \\
\citet{Cirasuolo07}      & K          &  0.25-2.25   &      blue triangle         \\
\hline
\hline
\end{tabular}
\end{center}
\label{lf_data}
\begin{flushleft}
\footnotesize {$^*$Central wavelengths corresponding to different wavebands 
are as follows: FUV=0.15$\mu$m, NUV=0.28$\mu$m, U=0.365$\mu$m, B=0.445$\mu$m, V=0.551$\mu$m, 
R=0.658$\mu$m, I=0.806$\mu$m, J=1.27$\mu$m, H=1.63$\mu$m and K=2.2$\mu$m.\\
$^\dagger$ These plotting symbols are used in Fig.~\ref{fig.fuv} and Fig.~\ref{fig.A1} 
for the $\rho_{\nu}$ obtained using different luminosity functions.} \\
\end{flushleft}
\label{table.main}
\end{table*}
%

We need to compute the galaxy emissivity, $\epsilon_{\nu,\rm G}(z)$, which is consistent with 
the observed luminosity functions of galaxies at different wavelengths and redshifts. The 
luminosity function, $\phi_{\nu}(L,z)$, observed at different $z$ and frequency $\nu$ is 
usually specified in the form of Schechter function. The comoving luminosity density, 
$\rho_{\nu}(z)$, for galaxies which is nothing but the space averaged comoving specific 
emissivity, 
\begin{equation*}
\rho_{\nu}(z)=\frac{\epsilon_{\nu,\rm G}(z)}{(1+z)^3} \,\, , 
\end{equation*} 
is given by an integral, 
%
\begin{equation}\label{rho}
\rho_{\nu}=\int_{L_{min}}^{\infty} {L \phi_{\nu}(L) dL}=\phi_{\nu}^{*} \,\, L^{*} \,\, \Gamma(\alpha+2, \,\,L_{min}/L^{*}).
\end{equation}
%
Here, $\phi_{\nu}^{*}$, $L^{*}$ and $\alpha$ are the Schechter parameters, $L_{min}$ is the 
luminosity corresponding to faintest galaxy at a redshift $z$ and $\Gamma$ is the incomplete 
gamma function. We dropped the subscript $z$ in above equation for clarity. The $\rho_{\nu}$ 
depends on the choice of $L_{min}$. In principle, one can always take $L_{min}=0$ for 
$\alpha >-2.0$ where the integral in Eq.~\ref{rho} converges. Generally, for galaxies at 
$z<2.5$, one finds $\alpha > -1.3$ \citep{Cucciati12}. In this case, the change in 
$\rho_{\nu}$, when one changes the $L_{min}$ from 0 to 0.01$L^{*}$, is less than $10\%$. We 
discuss the effect of adopting different the $L_{min}$ values in 
section~\ref{sec.res}.

The $\rho_{\nu_{0}}(z)$ measurements are used to determine the global star formation 
history \citep[see][]{Madau96, Lilly96} of the universe provided that the magnitude of the 
dust attenuation, $A_{\nu_{0}}(z)$, at any frequency $\nu_{0}$ and redshift $z$ is known. The 
average star formation rate density (SFRD), in units of M$_{\odot}$ yr$^{-1}$ Mpc$^{-3}$, is 
connected to $\rho_{\nu_{0}}$ through the relationship \citep{Kennicutt98},
%
\begin{equation}\label{Eq.conversion}
\text{SFRD}(z)=\zeta_{\nu_{0}} \times\rho_{\nu_{0}}(z)10^{0.4A_{\nu_{0}}(z)} .
\end{equation}
%
Here, $\zeta_{\nu_{0}} $ is a constant conversion factor which depends on $\nu_{0}$ and the 
initial mass function (IMF) assumed for galaxies. However, note that the relation given in 
Eq.~\ref{Eq.conversion} is an approximation as $\rho_{\nu_0}(z)$ can also have contributions 
from the old stellar population where the stars that are formed earlier are still shining at 
$z$.   

The derived SFRD($z$) and the SED produced from the instantaneous burst of star 
formation can be used to get the luminosity 
density at different $\nu$ and $z$. For an assumed initial mass function (IMF) and 
metallicity $Z$, the population synthesis models provide a SED in terms of specific 
luminosity, $l_{\nu}(\tau, Z)$, (in units of ergs s$^{-1}$ Hz$^{-1}$ per unit mass of stars 
formed) at different age $\tau$ of the stellar population. Since the timescales involved in 
the process of star formation (10$^5$ to 10$^7$ years) are relatively small, the SED from the 
instantaneous star burst can be directly convolved with the global SFRD($z$) to get the 
$\rho_{\nu}(z_0)$ by solving the following convolution integral 
\citep[see for eg,][HM12]{Kneiske02},
%
\begin{equation}\label{Eq.convolution}
\rho_{\nu} (z_0)=C_{\nu}(z_0)\int_{z_0}^{z_{max}}{\rm SFRD}(z)\,\,l_{\nu}[t(z_0)-t(z), Z]\,\,\frac{dt}{dz}\,\,dz\, ,
\end{equation}
%
where, the $\tau=t(z_0)-t(z)$ is an age of the stellar population at 
the redshift $z_{0}$ which went through an instantaneous burst of star formation at a 
redshift $z$ , $C_{\nu}(z_0)$ is the dust correction factor at $z_{0}$ and 
$\frac{dt}{dz}=[(1+z)H(z)]^{-1}$. 
The fact that the burst of star formations occurs at all epochs, $t$, but with the 
average star formation rates equals to the global SFRD$(t)$ is captured by the product 
of ${\rm SFRD}(z)$ and $l_{\nu}[t(z_0)-t(z), Z]$ in the convolution integral. We use 
$z_{max}=\infty$ as often used in the literature \citep[e.g.,][HM12]{Gilmore09, Inoue13}.
Later in section~\ref{sec.res}, we discuss the validity of the $z_{max}=\infty$ 
assumption and the effect of using different $z_{max}$. The dust correction factor, 
$C_{\nu}(z_0)$, for $\lambda <912$\AA~ is assumed to be equal to the escape fraction of 
hydrogen ionizing photons from galaxies as given in HM12. For $\lambda >912$\AA, we 
use $C_{\nu}(z_0)=10^{-0.4\,A_{\nu}(z_{0})}$, where, $A_{\nu}(z_{0})$ which is normalized at 
$\nu_{0}$ as given by,
\begin{equation}\label{Eq.k}
A_{\nu}(z_{0})=A_{\nu_{0}}(z_{0})\frac{k_{\nu}}{k_{\nu_0}}.
\end{equation}
%
Here, $k_{\nu}$ is a frequency dependent dust extinction curve. 

%
%
%
%
\begin{figure*}
\centering
  \includegraphics[bb=100 365 520 710,width=12cm,keepaspectratio,clip=true]{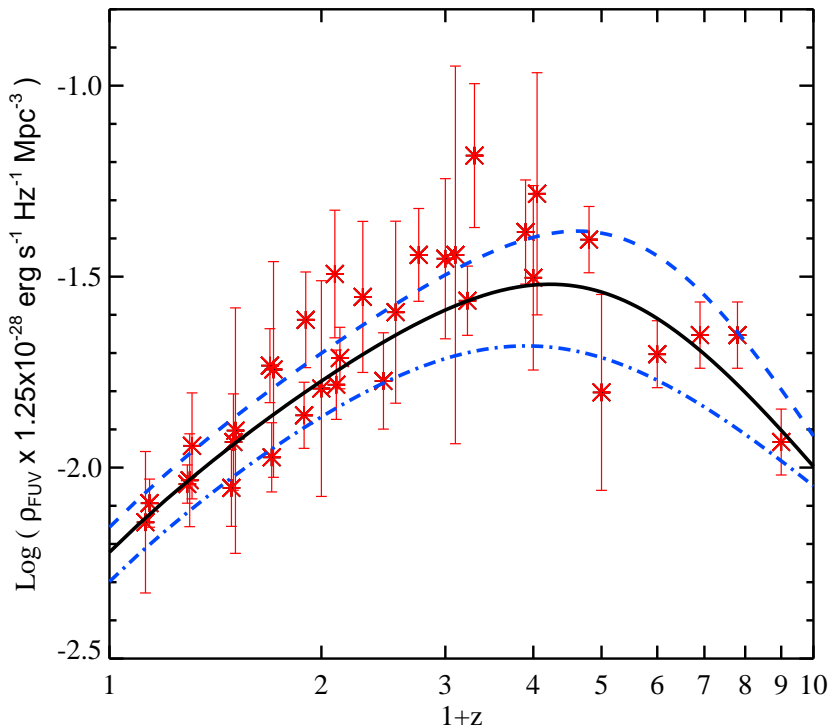}
  \caption{ The $\rho_{_{\rm FUV}}$ as a function of $z$ fitted with a functional form given in
Eq.~\ref{Eq.fit_form}. \emph{Solid, dashed} and \emph{dot-dash} 
curves are the median high and low fits, respectively. Data is taken from the references listed in Table~\ref{lf_data}
for the FUV band.} 
\label{fig.fit}
\end{figure*}
%
%
\section{Method to determine SFRD($z$) and A$_{\nu_0}(z)$}\label{sec.method}
In this sections, we summarize $\rho_{\nu}(z)$ measurements from the literature and the 
progressive fitting method which determines a unique combination of SFRD($z$) and 
A$_{\nu_0}(z)$ for an assumed extinction curve using $\rho_{\nu} (z)$. By construct, 
this combination of SFRD($z$) and A$_{\nu_0}(z)$ reproduces the emissivity measurements. 
%
%
\subsection{Compiled luminosity density measurements}
Motivated by the previous works of \citet{Stecker12} and \citet{Helgason12a}, we have 
compiled available observations of the galaxy luminosity functions and the corresponding 
$\rho_{\nu}$ at different rest wavelengths and $z$. In Table~\ref{lf_data}, we have 
given references along with the rest waveband and a redshift range over which the 
luminosity functions have been determined. In Table~\ref{lmin_table} in the Appendix, 
we list the faint end slopes of luminosity functions with the rest waveband and redshift 
along with the $L_{min}$ values we used to determine $\rho_{\nu}$. In general, we preferred 
the references where luminosity functions are determined in different wavebands the 
from the FUV (centered at $\lambda=0.15\mu$m) to K 
(2.2$\mu$m) band and with the largest possible coverage 
in redshift. This compilation has luminosity functions determined in the FUV band up to $z=8$, 
in the NUV and H band up to $z=3.5$ and for all other bands the measurements are available 
up to $z\sim2.5$. We take the $\rho_{\nu}(z)$ with the errors from the references where it is 
explicitly calculated. We use luminosity functions given in other references and compute 
$\rho_{\nu}$($z$) (using Eq.~\ref{rho}) with $L_{min}=0.01L^*$. Since there are more 
measurements of the $\rho_{\nu}$ in the FUV band and covering a large $z$ range, we choose 
$\nu_{0}=\nu_{\rm FUV}$, the frequency we use to determine SFRD($z$) as a frequency 
corresponding to the FUV band.
%
%
%
%
\subsection{Progressive fitting method}\label{sec.pfmethod}
We use a population synthesis model `{\sc Starburst99}'
\footnote{http://www.stsci.edu/science/starburst99/docs/default.html} \citep{Leitherer99},
to get the specific luminosity from stellar population of a typical galaxy, $l_{\nu}(t, Z)$, 
at an age $t$  and a metallicity $Z$ with an instantaneous burst of star formation. 
In these simulations, we consider a constant metallicity of $Z=0.008$ over all $z$. 
Later in section~\ref{sec.res}, we also discuss the effect of using different values of 
metallicity. We use the Salpeter IMF with the exponent of 2.35 and the stellar mass range 
from 0.1 to 100 M$_{\odot}$. For this particular galaxy model, we find the conversion factor 
for connecting $\rho_{FUV}(z)$ and SFRD($z$) (see Eq.~\ref{Eq.conversion}) to be 
$\zeta_{\nu_0}=1.25\times10^{-28}$. As described before, the reference frequency, 
$\nu_0$ which we use corresponds to the frequency of the FUV band. Note that, this conversion 
factor  1.25$\times10^{-28}$  is 11\% smaller than widely used, 1.4$\times10^{-28}$, quoted 
by \citet{Kennicutt98}. This difference is mainly because of the updated population synthesis 
model and the assumed metallicity. 

We fit a functional form  to the compiled $\rho_{\rm FUV}$ data and obtained its parameter
using the {\sc mpfit IDL} routine \footnote{{\sc mpfit} is a robust non-linear least square 
fitting IDL program used to fit model parameters for a given data \citep{mpfit}.} 
that uses $\chi^2$ minimization. At high redshifts, we take 20\% errors on the 
$\rho_{\rm FUV}$ calculated from the luminosity function given by \citet{Bouwens11}. 
We convert the asymmetric errors into symmetric errors by taking the average of them. 
To fit $\rho_{\rm FUV}(z)$, we use a following functional form that was originally used by 
\citet{Cole01} to fit the SFRD($z$), 
\begin{equation}\label{Eq.fit_form}
\rho_{\rm FUV}(z)=\frac{a+bz}{1+(z/c)^{d}}.
\end{equation}
There is a large scatter in the $\rho_{\rm FUV}(z)$ data. Therefore, along with this fit 
(hereafter, median fit) we construct 1-$\sigma$ upper and lower limit fits (hereafter we 
refer to them as the high and low $\rho_{\rm FUV}$ fits, respectively). These $\rho_{\rm FUV}$ 
fits multiplied by $1.25\times10^{-28}$ are nothing but the different SFRD($z$) with 
the $A_{\text{FUV}}(z)=0$ (see Eq.~\ref{Eq.conversion}) which are plotted in 
Fig.~\ref{fig.fit}. The values of fitting parameters for the median 
$\rho_{\rm FUV}\times1.25\times10^{-28}$ fit are $a=(6\pm1)\times 10^{-2}$, 
$b=(11\pm2)\times 10^{-2}$, $c=4.41\pm0.58$ and $d=3.15\pm0.62$. 
We construct 1-$\sigma$ high and low  $\rho_{\rm FUV}$ fits by adding and subtracting 
the error in each parameter from its best fit values, respectively (see, Fig.~\ref{fig.fit}).
We determine the combinations of  SFRD($z$) and  $A_{\rm FUV}(z)$  for all three (low, median 
and high) $\rho_{\rm FUV}$ fits with different extinction curves as discussed below.
%
%
\begin{figure*}
\centering
  \includegraphics[bb=100 365 520 710,width=12cm,keepaspectratio,clip=true]{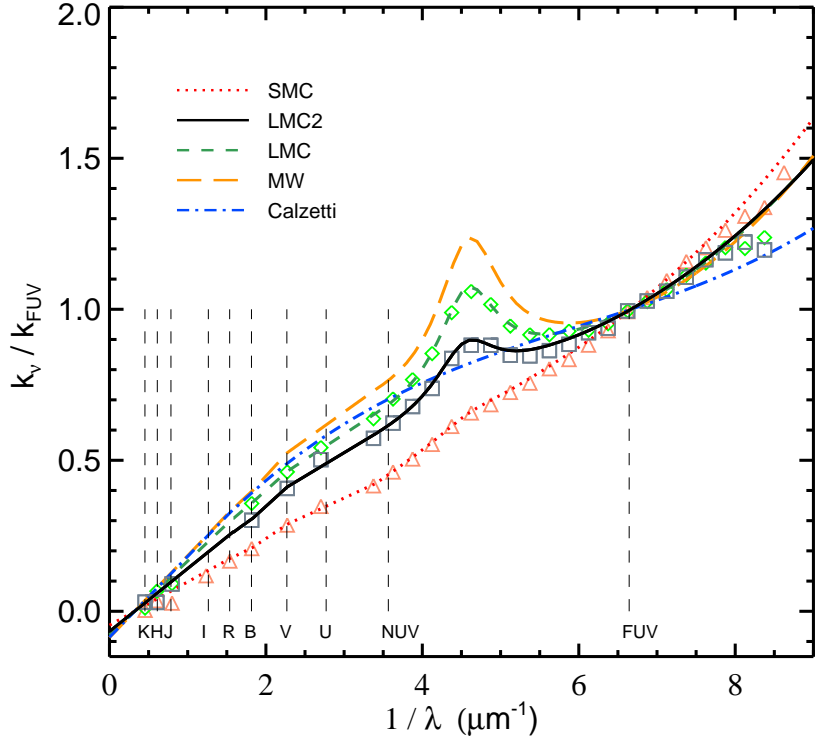}
  \caption{ Extinction curves normalized at the FUV band for SMC, LMC, LMC Supershell (LMC2), 
Milky-Way (MW) and nearby starburst galaxies by \citet{Calzetti}. Here, \emph{triangles, 
squares} and \emph{diamonds} represent the mean extinction curve measurements from 
\citet{Gordon03} normalized at the FUV band for the SMC, LMC2 and LMC, respectively. 
Different wavebands are marked with the \emph{vertical dashed} lines to show the difference 
in different extinction curves at those wavelengths. }
\label{fig.k}
\end{figure*}
%
%

The average extinction curve for high redshift galaxies is one of the key unknowns in the
astronomy. However, the mean extinction curves for our galaxy, Small and Large Megallenic 
Clouds (SMC and LMC) and some low redshift starburst galaxies are well known 
\citep{Lequeux82,Clayton85,Calzetti94}. It is a general practice to use the average 
extinction curve determined for the nearby starburst galaxies by \citet{Calzetti}
\footnote{Note that, sometimes it is also called as an attenuation curve or an 
obscuration curve \citep{Calzetti01}. However, in this paper, along with other four 
extinction curves we call it as an extinction curve for the uniformity.} 
for the high redshift galaxies. Here, along with 
\citet{Calzetti} extinction curve, we use extinction curves determined for SMC, LMC and 
LMC supershell (LMC2) from \citet{Gordon03} and  Milky-Way (MW) from \citet{Misselt99}. 
In particular, this set of extinction curves encompasses a wide range of dust properties 
typically present in the astronomical domain. Since, we are using $\rho_{\rm FUV}$ 
measurements for determining the SFRD($z$), we normalize all the extinction curves $k_{\nu}$ 
at $\nu$ corresponding to the FUV band ($0.15\mu$m). In Fig.~\ref{fig.k}, we have plotted 
the $k_{\nu}/k_{\text{FUV}}$ for different extinction curves as a function of $\lambda^{-1}$ 
along with the respective measured data points from \citet{Gordon03} for the SMC, the LMC and 
the LMC2. In Fig.~\ref{fig.k}, we also mark the different $\lambda^{-1}$ for the wavebands at 
which we have compiled the $\rho_{\nu}$ measurements to determine the $A_{\rm FUV}$ and 
the SFRD. 

From Eq.~\ref{Eq.conversion} it is clear that the SFRD($z$) and $A_{\text{FUV}}(z)$ are 
degenerate quantities and different combinations of them can give the same 
$\rho_{\text {FUV}}$. However, the measured $\rho_{\nu}$ values at different frequencies other 
than the FUV band along with the assumed extinction curve break this degeneracy. Here we 
introduce a novel method that, by using the multi-wavelength and multi-epoch luminosity 
functions, determines the $A_{\rm FUV}(z)$ and SFRD($z$) uniquely for an assumed extinction 
curve. In this method we initially fix the $A_{\rm FUV}$ and SFRD at some higher redshifts and 
then using this we progressively determine $A_{\rm FUV}$ and SFRD at lower redshifts. This 
`progressive fitting method' is described below in details.

Combining Eq.~\ref{Eq.conversion}, \ref{Eq.convolution} and \ref{Eq.k}, the 
$\rho_{\nu}$($z_0$) can be written as, 
%
\begin{eqnarray}
\lefteqn{\rho_{\nu} (z_0)= 1.25\times10^{-28}\times 10^{\big[-A_{\rm FUV}(z_{0})\frac{k_{\nu}}{k_{\rm FUV}}\big]}}\nonumber \\ 
& & \times \int_{z_0}^{\infty}\rho_{\rm FUV}(z)\; 10^{0.4A_{\rm FUV}(z)} \, l_{\nu}[t(z_0)-t(z), Z]\,\,\frac{dt}{dz}\,\,dz\,.
\label{Eq.tau}
\end{eqnarray}
%
For a given extinction curve, $k_{\nu}$, and our $\rho_{\rm FUV}(z)$ fits, the only unknown 
in the above equation is $A_{\rm FUV}(z)$ for $z\ge z_0$. Therefore, to get the 
$\rho_{\nu}$($z_0$) one needs to know the $A_{\rm FUV}(z)$ for all $z\ge z_0$. The procedure 
we followed to get the $A_{\rm FUV}(z)$ for each extinction curve $k_{\nu}$ using the 
$\rho_{\nu}(z)$ measurements is given below:
\begin{enumerate}
  \item We choose the highest possible redshift $z_{th}$ where we have 
$\rho_{\nu} (z_{th})$ measurements in most of the wavebands. 
  \item For all $z\ge z_{th}$, we assume a functional form for $A_{\rm FUV}(z)$. 
  \item We fix the normalization of this function and hence the value of 
$A_{\rm FUV}(z_{th})$ by matching the predicted $\rho_{\nu}(z_{th})$ with the measured ones 
at different wavebands (other than the FUV band) using the least square minimization. This 
fixes the $A_{\rm FUV}(z)$ for $z\ge z_{th}$. Then we call $z_{th}$ as $z_1$.
  \item We choose the next redshift $z_0<z_1$ which is the next nearby lower redshift 
where we have multi-wavelength $\rho_{\nu}(z_0)$ measurements.  
  \item We assume $A_{\rm FUV}(z)$ is constant and equal to $A_{\rm FUV}(z_0)$ in 
between the redshifts $z_0$ and $z_1$. For $z\ge z_1$ we use the $A_{\rm FUV}(z)$ as 
determined earlier. Then we calculate the $\rho_{\nu}(z_0)$ for different values of 
$A_{\rm FUV}(z_0)$.
  \item We compare the resultant $\rho_{\nu}(z_0)$ with the measured one at different 
wavebands and determine the best fit $A_{\rm FUV}(z_0)$ by the least square minimization. 
This fixes the $A_{\rm FUV}(z)$ for $z\ge z_0$. Then we call this $z_0$ as $z_1$.
  \item We repeat the steps 4 to 6 until we reach the lowest $z$ where we have 
multi-wavelength $\rho_{\nu}(z)$ measurements. This provides us the best fit values of 
$A_{\rm FUV}(z)$ and SFRD($z$) over the whole redshift range for a given extinction 
curve.
\end{enumerate}
%
%
\begin{figure*}
\centering
  \includegraphics[bb=68 360 540 690,width=12.6cm,keepaspectratio,clip=true]{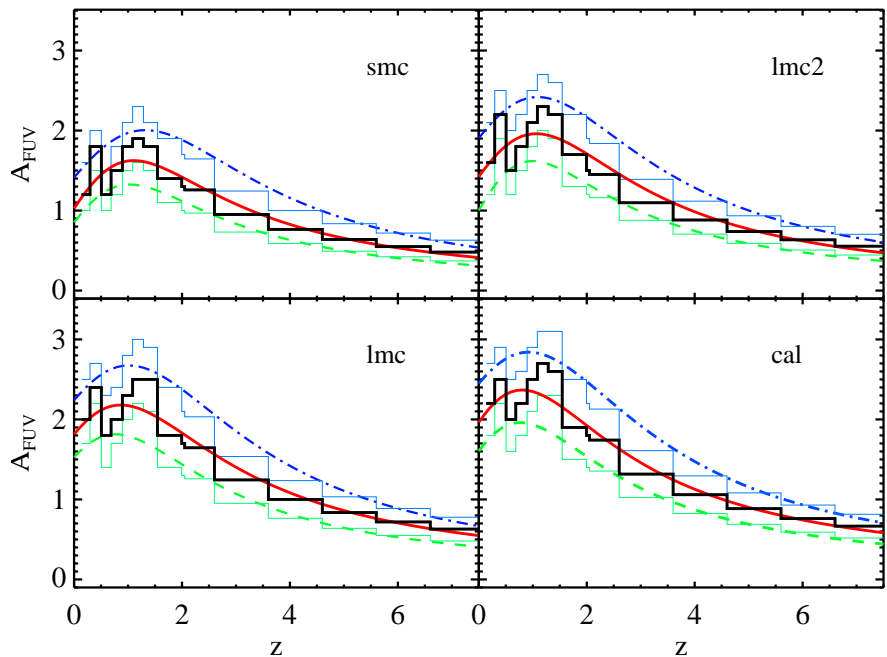}
  \caption{ The $A_{\rm FUV}$ as a function of $z$. \emph{Solid black histogram} is a best 
fit $A_{\rm FUV}$ determined by the method described in Section 5.1 for different extinction 
curves appropriately labeled in each panel. \emph{Solid blue and green histograms} are the 
best fit $A_{\rm FUV}$ determined using low and high $\rho_{\rm FUV}$ fits. The 
$A_{\rm FUV}(z)$ is fitted with a functional form given in Eq.~\ref{Eq.fit_form}. 
\emph{Solid red, dashed green} and \emph{dot-dash blue} curves are fit to the $A_{\rm FUV}$ 
obtained using  median, high and low $\rho_{\rm FUV}$ fits, respectively. 
The fitting parameters are given in Table~\ref{dust_parm}.}
\label{fig.dust_fit}
\end{figure*}
%
%

In Fig.~\ref{fig.dust_fit} we show the $A_{\rm FUV}(z)$ obtained (\emph {histograms}) 
using the progressive fitting method described above for different extinction curves.
We fit a continuous function through the resultant $A_{\rm FUV}(z)$ using a functional form 
same as the one we used to fit $\rho_{\rm FUV}(z)$ measurements (given in 
Eq.~\ref{Eq.fit_form}). For fitting this functional form we use {\sc mpfit idl} routine 
by taking 10\% errors for all. To demonstrate the procedure described here, in 
Fig.~\ref{fig.dust_fit}, we also show the resultant $A_{\rm FUV}$($z$) obtained using the 
high, low and median $\rho_{\rm FUV}$ fits (\emph{histograms}) along with its fitted 
functional form for different extinction curves. Since, we show Fig.~\ref{fig.dust_fit} for 
the purpose of demonstrating our `progressive fitting method', for clarity, we do not show 
$A_{\rm FUV}(z)$ obtained for Milky-Way extinction curve. Note that this resultant 
$A_{\rm FUV}(z)$ will directly give the corresponding SFRD$(z)$ (see Eq.~\ref{Eq.conversion}). 
We also fit SFRD$(z)$ using the same functional form (see Eq.~\ref{Eq.fit_form}).

Our aim is to get the combinations of $A_{\rm FUV}$($z$) and SFRD($z$) which will reproduce 
the measured $\rho_{\nu}(z)$ obtained using the observed luminosity functions at different 
wavebands and different $z$. The $\rho_{\nu}(z)$ measurements are taken from different 
references and they have different biases and error estimates. Therefore, to minimize the 
uncertainty and determine the $A_{\rm FUV}$ over large $z$ range uniquely, we have to choose 
$\rho_{\nu}(z)$ measurements which span many wavebands and large $z$ range and possibly 
reported by the same group so that the effect of various biases will be minimum. Fortunately 
this requirement is satisfied by the  $\rho_{\nu}$ measurements reported in \citet{Tresse07} 
where the $\rho_{\nu}$ is measured over seven different wavebands (from FUV to I band)
and at the same redshift bins spanning up to $z=2$. Therefore, to get a robust 
$A_{\rm FUV}(z)$ and SFRD($z$) combination we choose the observed $\rho_{\nu}(z)$ given by 
\citet{Tresse07} and take $z_{th}=2$. We assume that the form of the $A_{\rm FUV}$($z$) for 
$z \ge 2$ goes as $1/(1+z)$ and independent of the extinction curve used. We show later that 
this assumed form gives the $A_{\rm FUV}$($z$) consistent with other independent measurements. 
This trend of decreasing $A_{\text{FUV}}$ at higher $z$ has been previously observed  
\citep[see for e.g][]{Burgarella13, Cucciati12, Bouwens09, Takeuchi05}. 
This is consistent with the picture of gradual build up of dust in galaxies with cosmic time 
as evident from the fact that galaxies at very high redshifts ($z>5$) are bluer than
the $z\sim2$ to $4$ galaxies \citep{Bouwens09}. 

We calculate the $A_{\rm FUV}(z)$ and corresponding SFRD($z$) for all the five extinction 
curves used in this paper using the low, high and median $\rho_{\rm FUV}(z)$ fits. As we show 
later, we use the $\rho_{\nu}(z)$ obtained using the combinations of $A_{\rm FUV}(z)$ and 
SFRD($z$) to estimate the $\rho_{\nu}(z)$ at FIR wavelengths and the EBL at different 
redshifts. We denote the obtained combinations of $A_{\rm FUV}(z)$ and SFRD($z$), the 
$\rho_{\nu}(z)$ and the EBL using different extinction curves as the `smc', `lmc', `lmc2', 
`mw' and `cal' models based on the SMC, LMC, LMC2, Milky-Way and \citet{Calzetti} extinction 
curves used, respectively. For most comparisons we use our default models which are obtained 
using median fits through $\rho_{\rm FUV}$ points. We use the predictions of the high and the 
low fits only when we discuss the spread. For clarity in the subsequent discussions, whenever 
we use the `high (low) model' we mean the relevant quantity (like $\rho_{\nu}$, $A_{\rm FUV}$, 
SFRD and EBL) obtained with the high (low) $\rho_{\rm FUV}$ fit and the `model' extinction 
curve. When we denote only `model' we mean the relevant quantity obtained using the median 
$\rho_{\rm FUV}$ fit and that `model' extinction curve.

In the following section we discuss the resultant $\rho_{\nu}(z)$, $A_{\text{FUV}}(z)$ and 
SFRD(z) determined using the method described in this section.
%
%
%
\section{Dust attenuation and star formation history}
\label{sec.sfr}
\subsection{Reproducing $\rho_{\nu}(z)$ measurements}
%
\begin{figure*}
\centering
   \includegraphics[bb=10 110 575 780, width=16cm, keepaspectratio,clip=true]{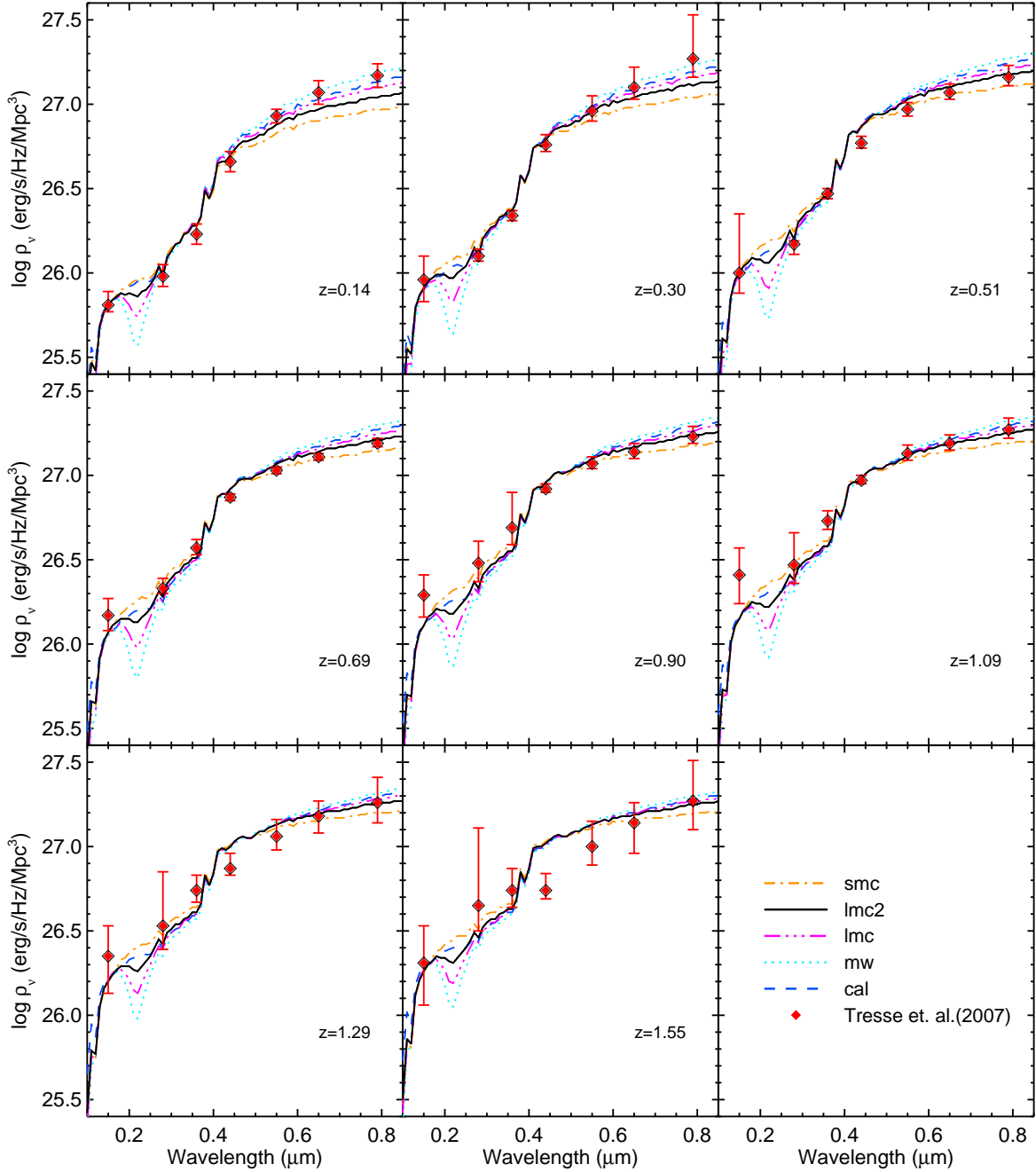}
  \caption{ \small{The average comoving galaxy emissivity, $\rho_{\nu}$, at different $z$ 
calculated using our best fit combination of the SFRD($z$) and $A_{\rm FUV}(z)$ obtained 
for different models (with median $\rho_{\rm FUV}$ fits).The red diamonds are the $\rho_{\nu}$ 
measurements from \citet{Tresse07}. (See Fig.~\ref{fig.A1} in Appendix for $\rho_{\nu}$($z$) 
obtained for the low and high models at different wavebands along with the compiled 
luminosity density measurements.)}}
\label{fig.zband}
\end{figure*}
%
\begin{figure}
\centering
    \includegraphics[bb=100 365 520 715,width=9cm,keepaspectratio,clip=true]{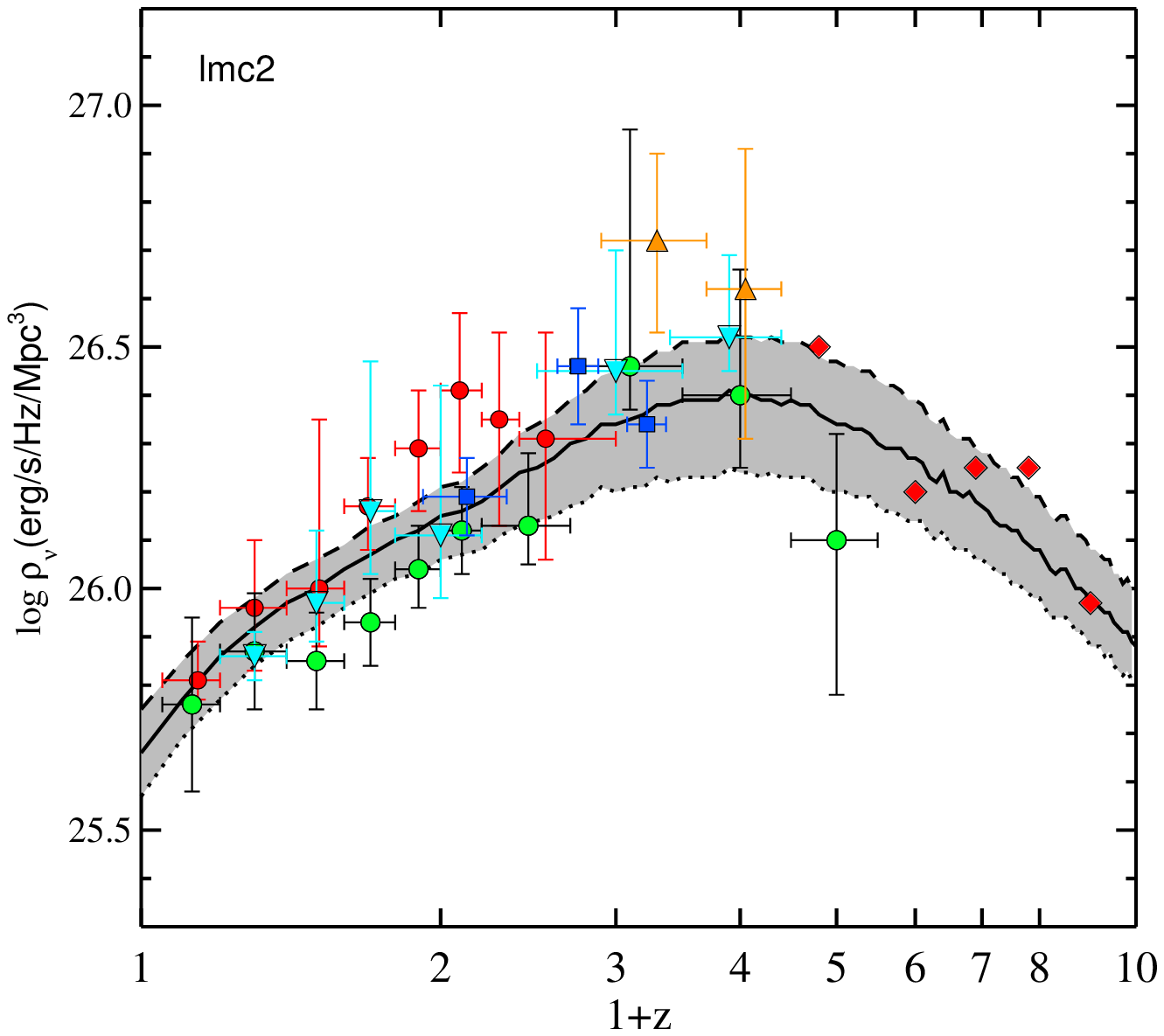}
\caption{ The FUV band comoving luminosity density with $z$. Solid, dashed and dotted lines 
represent  $\rho_{\nu}$ calculated at the FUV band using the median, high and low `lmc2' 
model, respectively. The plotting symbols and corresponding references are mentioned in
Table~\ref{lf_data} for the FUV  band. The $\rho_{\rm FUV}$ calculated for different models 
have negligible difference with respect to each other. Therefore, as a representative for all 
other models, we show $\rho_{\rm FUV}$ only for the `lmc2' model.}
\label{fig.fuv}
\end{figure}
%
In Fig.~\ref{fig.zband}, we plot the $\rho_{\nu}$ obtained using convolution integral 
(Eq.\ref{Eq.convolution}) for the best fit combinations of SFRD and $A_{\rm FUV}$ at different 
$z$ along with the measurements of \citet{Tresse07}. Note that, to get the $A_{\rm FUV}$ by 
least square minimization we use the $\rho_{\nu}$ measurements of \citet{Tresse07} in all 
wavebands except at the FUV band. However, these $\rho_{\rm FUV}$ measurements along with 
many other reported in the literature up to $z=8$ (see Table~\ref{lf_data} and 
Table~\ref{lmin_table}) goes into fitting the $\rho_{\rm FUV}$ (as shown in 
Fig.~\ref{fig.fit}). All our five models show very a good agreement with the measurements 
of \citet{Tresse07} in all wavebands (including the FUV band). The difference in the 
strength of 2175\AA~absorption feature arises because of using different extinction curves.

In Fig.~\ref{fig.fuv}, along with the compiled measurements, we plot the $\rho_{\nu}$ at the 
FUV band obtained by using the combination of the SFRD$(z)$ and the $A_{\text{FUV}}(z)$ for 
the low, median and high `lmc2' model. There are negligible differences in the 
$\rho_{\rm FUV}$ obtained for different models. Therefore, for clarity, we do not show the 
similar $\rho_{\rm FUV}$ plots for other models. 

For the high $z$ and all other wavebands, we show our estimated $\rho_{\nu}$ along with the 
compiled measurements in Fig.~\ref{fig.A1} in the appendix (see Table~\ref{lf_data} for 
the references and plotting symbols). Even though we use measurements of $\rho_{\nu}$ up to 
$z\sim2$ and up to wavelength corresponding to the I band to get the $A_{\rm FUV}(z)$, our 
estimated $\rho_{\nu}$ matches well with various measurements up to $z\sim 4$ from the NUV 
to K band. This implies that our determined combinations of the  $A_{\rm FUV}(z)$ and the 
SFRD($z$) are valid over a large $z$ range and suggests that our assumption of decreasing 
dust attenuation at high $z$ is also valid. However, note that, at the high redshifts (i.e, 
$z>4$) there are no measurements of $\rho_{\nu}$ except at the FUV band. In the H band, our 
calculated $\rho_{\nu}$($z$) is slightly over-estimated than the measured ones. However, as 
there are very few measurements we do not attempt to address this disagreement. 

The good matching between the observations and the model predictions suggests that we have a 
consistent combination of the $A_{\rm FUV}(z)$ and SFRD($z$) for each extinction curve under 
consideration. The evolution of our best fit $A_{\rm FUV}$ and the corresponding SFRD with 
$z$ is discussed in the next section.
%
%
%
\subsection{Redshift evolution of $A_{\text{FUV}}$}\label{sec.dust}
Understanding the dust attenuation and its wavelength and redshift dependences are very 
important to derive the intrinsic SFRD($z$) accurately. Dust attenuation is measured by using 
either of the SED fitting techniques, the Balmer decrement method or by comparing the FUV and 
the IR luminosity function measurements. It has also been noticed that at any given 
$z$, the derived $A_{\text{FUV}}$ may also depend on the galaxy luminosity and the stellar 
mass of the galaxy \citep[see for e.g.,][]{Bouwens12}. Recently it has been shown 
that the shape of the extinction curve strongly depends on the distribution of the dust in 
the galaxies and the viewing geometries where scattering plays an important role 
\citep{Chevallard13}. As our main purpose is to calculate the EBL, we are mainly 
interested in the volume averaged star formation rates and emissivity. Therefore, to 
calculate the average dust correction as a function of $z$, for simplicity we do 
not consider the dependence of $A_{\text{FUV}}$ on galaxy luminosity or stellar mass 
and the dependence of $k_{\nu}$ on scattering and viewing geometries. In this section, 
we compare the $A_{\text{FUV}}(z)$ obtained for different extinction curves with the 
$A_{\text{FUV}}$ measurements in the literature based on other independent approaches.
%
%
\begin{figure}
\centering
   \includegraphics[bb=140 20 520 770,width=9.5cm, keepaspectratio,clip=true]{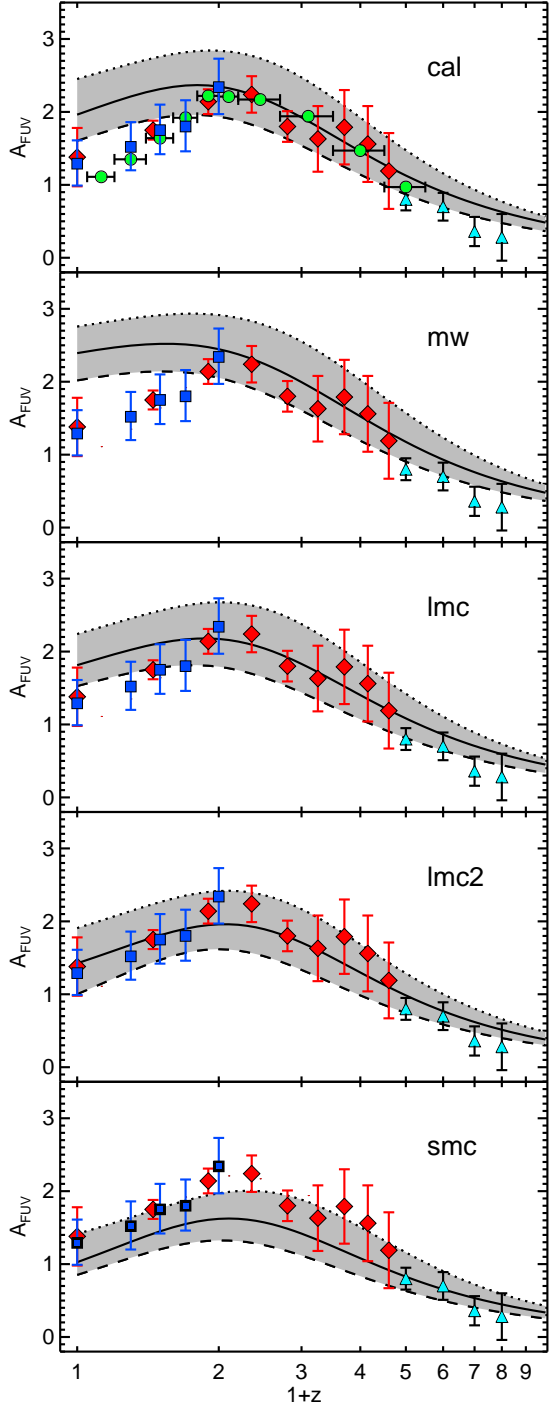}
  \caption{ \small{ Our best fit dust attenuation, $A_{\rm FUV}$, in magnitude as a 
function of redshift calculated by using
different extinction curves. \emph{Dotted, solid} and  
\emph{dashed} lines represent values of the best fit $A_{\rm FUV}$
obtained using the low, median and high models, respectively. 
\emph{Green circles} represent the A$_{\rm FUV}$ determined 
through the SED fitting by \citet{Cucciati12} using Calzetti extinction 
curves.  \emph{Red diamonds} and  \emph{Blue squares} represents 
the A$_{FUV}$ measured through $\rho_{\text{FIR}}$ 
to $\rho_{\text{FUV}}$ ratio by \citet{Burgarella13} and 
\citet{Takeuchi05}, respectively.  
\emph{Cyan triangles} are from \citet{Bouwens12}.}
}
\label{fig.dust_shade_all}
\end{figure}
%
%
%
%
%
\begin{figure}
\centering
   \includegraphics[bb=140 20 520 770,width=9.5cm, keepaspectratio,clip=true]{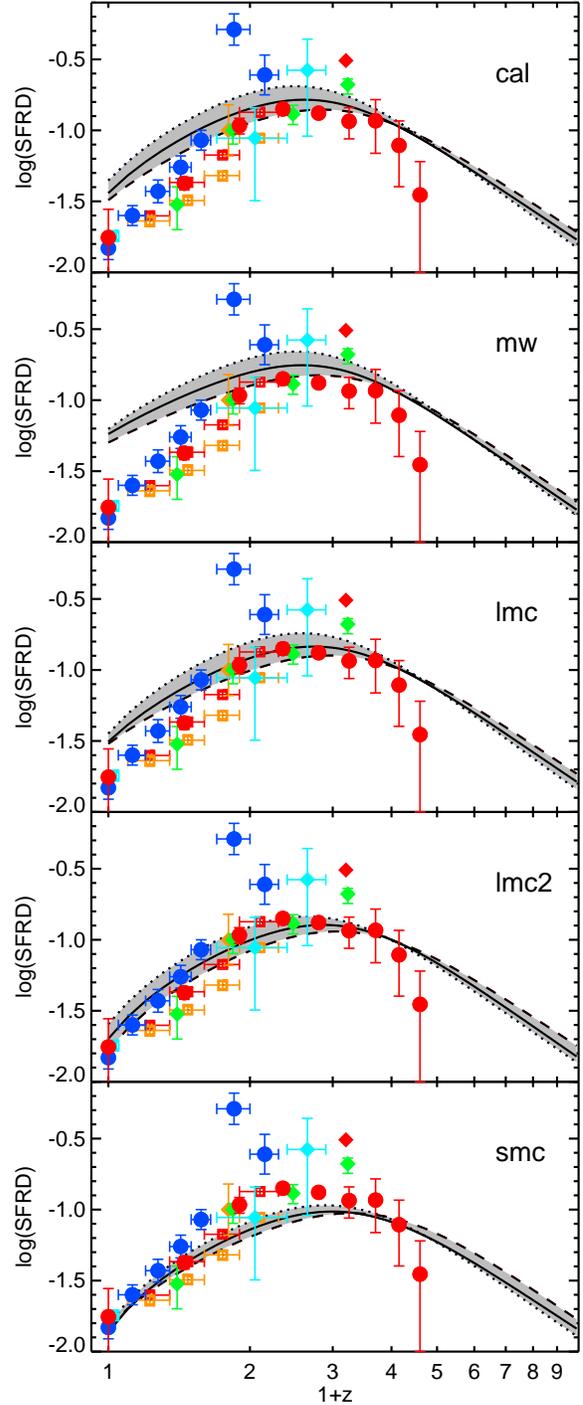}
  \caption{ \small{ Our best fit SFRD($z$) in units  
M$_{\odot}$ yr$^{-1}$ Mpc$^{-3}$ obtained using different extinction curves. 
\emph{Dotted}, \emph{solid} and \emph{dashed} lines represent 
values of best fit SFRD($z$) obtained using the low, median and 
high models, respectively. Here, \emph{squares}, 
\emph{diamonds} and \emph{circles} represent the SFRD($z$) determined from  
the radio, the H-$\alpha$ and the FIR observations, respectively. 
References and plotting symbols used here are provided in Table~\ref{sfr_data}. 
The SFRD($z$) obtained using the LMC2 extinction curve shows good agreement with the
different dust independent measurements.}}
\label{fig.sfrd_shade_all}
\end{figure}
%
%
\begin{table}
\caption{Fitting parameters for the $A_{\rm FUV}$$^*$}
\begin{tabular}{ l l c c c c c c}
\hline              
\hline
Extinction curve & $\rho_{FUV}$ fit$^\dagger$   & a & b & c & d &  \\
\hline
\hline
 SMC            &   Low       &  1.41 &    0.79 &    2.51 &    2.32  \\ 
                &   Median    &  1.03 &    1.01 &    1.87 &    2.16  \\ 
                &   High      &  0.85 &    0.86 &    1.77 &    2.16  \\
\hline
 LMC2           &   Low       &  1.91 &    0.85 &    2.40 &    2.23  \\ 
                &   Median    &  1.42 &    0.93 &    2.08 &    2.20  \\ 
                &   High      &  1.00 &    1.16 &    1.66 &    2.14  \\
\hline              
 LMC            &   Low       &  2.24 &    0.79 &    2.51 &    2.21  \\ 
                &   Median    &  1.81 &    0.82 &    2.12 &    2.06  \\ 
                &   High      &  1.53 &    0.73 &    1.93 &    2.05  \\
\hline              
 Milky-Way      &   Low       &  2.76 &    0.44 &    2.98 &    2.14  \\ 
                &   Median    &  2.39 &    0.48 &    2.44 &    1.97  \\ 
                &   High      &  2.02 &    0.49 &    2.15 &    1.97  \\
\hline              
 Calzetti       &   Low       &  2.45 &    0.79 &    2.48 &    2.15  \\ 
                &   Median    &  1.96 &    1.00 &    1.94 &    2.02  \\ 
                &   High      &  1.61 &    0.91 &    1.81 &    2.03  \\                                                                               
\hline              
\hline
\end{tabular}
\hfill
\label{dust_parm}
\begin{flushleft}
\footnotesize {*The fitting form is $A_{\rm FUV}(z)=\frac{a+bz}{1+(z/c)^{d}}$.}\\
\footnotesize {$^\dagger$Note that, the models with low (high) $\rho_{\rm FUV}$ fit give higher (lower) 
A$_{\rm FUV}$ than the median model as explained in the text.}\\
\end{flushleft}
\end{table}
%

The fitting parameters for $A_{\text{FUV}}(z)$ for different extinction curves are given in 
Table~\ref{dust_parm}. In Fig.~\ref{fig.dust_shade_all}, we plot the range of 
$A_{\text{FUV}}(z)$ for different extinction curves along with the measurements of 
\citet{Takeuchi05}, \citet{Cucciati12}, \citet{Burgarella13} and \citet{Bouwens12}. 
\citet{Takeuchi05} and \citet{Burgarella13} determined the $A_{\text{FUV}}$ using the ratio of 
the FUV to FIR band luminosity density. \citet{Cucciati12} have calculated the $A_{\rm FUV}$ 
using the \citet{Calzetti} extinction curve and used the SED fitting technique. At very high 
redshifts \citet{Bouwens12} determined the effective dust extinction using the UV-continuum 
slope $\beta$ distribution and the IRX-$\beta$ relationship \citep[see,][]{Meurer99}. 
We take the effective extinction calculated for the luminosity function integrated up to -17.7 
magnitude from  \citet{Bouwens12} (from table 6 of their paper). 
 
The shaded region in Fig.~\ref{fig.dust_shade_all} is obtained by using the low, high and 
median $\rho_{\rm FUV}$ fits to determine the $A_{\rm FUV}$. Since the SFRD is directly 
related to the $\rho_{\rm FUV}$ and $A_{\rm FUV}$, when we use the low (high) $\rho_{\rm FUV}$ 
fits, to get the same $\rho_{\nu}$ at different wavebands we need higher (lower) SFRD and 
hence higher (lower) $A_{\rm FUV}$. In other words, the low (high) $\rho_{\rm FUV}$ implies 
that the galaxies are more red (blue) which suggest that these galaxies should have more 
(less) dust extinction. This trend is evident from Fig.~\ref{fig.dust_shade_all}, where the 
dotted and dashed curves show the $A_{\rm FUV}$ obtained using the low and the high 
$\rho_{\rm FUV}$ fits, respectively. 

The shaded region in Fig.~\ref{fig.dust_shade_all} represents the allowed range of 
$A_{\rm FUV}$. For each  assumed extinction curve, we get a different allowed range 
for the $A_{\rm FUV}$ and the difference is prominent at redshifts $z<1$. For redshifts 
$1<z<2$, we find that the $A_{\rm FUV}$ values remain constant or show a mild decrease with 
increase in $z$. At high redshifts, i.e $z>2$, our assumption of decreasing dust attenuation 
plays a role in getting similar allowed range of the $A_{\rm FUV}$ for all assumed extinction 
curves. As can be seen from Fig.~\ref{fig.dust_shade_all}, the allowed $A_{\rm FUV}$ range for 
$z>2$ nicely follows that of other independent measurements rendering support to our 
assumption. Apart from the $A_{\rm FUV}$ determined for the `mw' model, for all the other 
models we find a moderate increase in the $A_{\rm FUV}$ with redshift up to $z=1$ from $z=0$. 
This trend of increasing FUV band dust attenuation magnitude has been detected 
previously \citep[see,][]{Takeuchi05, Cucciati12, Burgarella13} as shown in 
Fig.~\ref{fig.dust_shade_all}. However, at $z\le0.8$, our estimated $A_{\text{FUV}}(z)$  
for `cal', `mw', and `lmc' models are higher than these measurements. The $A_{\rm FUV}$ 
determined for `smc' model matches well in all redshifts expect that it under-predicts 
$A_{\rm FUV}$ at $1<z<2$. Overall, a good match with these measurements of the $A_{\rm FUV}$ 
is obtained over the large $z$ range for the `lmc2' model.

From the very good agreement between the $A_{\rm FUV}(z)$ determined for the `lmc2' model and 
the measurements of \citet{Burgarella13} and \citet{Takeuchi05}, we conclude that the 
average extinction curve which is applicable for galaxies over wide range of redshifts is 
most likely to be similar to LMC2 extinction curve.

Recently, \citet{Kriek13} using SED of the galaxies investigated the dust extinction curves 
for 32 different spectral classes of galaxies over $0.5 \le z \le2$. They found that the 
Milky-Way and Calzetti extinction curves provide poor fits to the UV wavelengths for all SEDs. 
They concluded that the SED with the $2175\AA$ UV bump albeit weaker in strength compared to 
the Milky-Way is preferred. \citet{Buat12} studied a sample of 751 galaxies with redshift 
$0.95 < z < 2.2$ and found that the mean parameters describing the dust attenuation curves 
are similar to those found for the LMC2 extinction curve. This is consistent with what we find 
here. It is interesting to note that in the case of intervening absorption systems seen in the 
QSO spectra, the high percentages of systems with the $2175\AA$ absorption feature detection 
favors the LMC2 extinction curve \citep[see for e.g.][]{Srianand08, Noterdaeme09, Jiang13}. 
It has also been observed that the extinction curves for individual galaxies depend on 
the type and other galaxy properties \citep{Wild11, Chevallard13, Kriek13}. Therefore, single 
universal extinction curve for all galaxies may not be realistic. However, our study suggests 
that for estimating the volume average properties like the SFRD, $A_{\rm FUV}$ and EBL the 
LMC2 extinction curve should be preferred. 
%
\subsection{Redshift evolution of SFRD}\label{sec.sfrd}
We assume that the SFRD is a smooth and continuous function of $z$ and fit it with the 
same functional form (using Eq.~\ref{Eq.fit_form}) we are using to fit the $\rho_{\rm FUV}$ 
and $A_{\rm FUV}$. The fitting parameters for the SFRD$(z)$ are given in Table~\ref{sfrd_parm}.
In Fig.~\ref{fig.sfrd_shade_all}, we plot the SFRD($z$) for all the five extinction curves 
with their high and low models. As explained earlier we get the high (low) SFRD for low 
(high) $\rho_{\rm FUV}$ fits. Since the SFRD is directly proportional to the $\rho_{\rm FUV}$ 
and $A_{\rm FUV}$, at $z>3$, where differences in the  $A_{\rm FUV}$ for all high and low 
models are small, the term $\rho_{\rm FUV}$ dominates and the SFRD($z$) curves cross each 
other as demonstrated in Fig.~\ref{fig.sfrd_shade_all}.

\begin{table*}
\caption{SFRD from different observations}
\begin{center}
\begin{tabular}{l c c c c c c }             
\hline
\hline
 Reference & Technique & Redshift range &  Plotting Symbols   \\
\hline
\cite{Shim09}           & H-$\alpha$     & 0.7-1.9     &   cyan diamond            \\
\cite{Tadaki11}         & H-$\alpha$     & 2.2         &   red diamond              \\
\cite{Sobral13}         & H-$\alpha$     & 0.4-2.3     &   green  diamond            \\
\cite{Ly11}             & H-$\alpha$     & 0.8         &   orange diamond            \\
\cite{Condon02}         & 1.4 GHz        & 0.02        &      cyan squares        \\
\citet{Smolcic09}       & 1.4 GHz        & 0.1-1.3     &   red \& orange squares   \\
\cite{Rujopokarn10}     & FIR            & 0-1.3       &   blue circles            \\
\cite{Burgarella13}     & FIR            & 0-4         &      red circles          \\
\hline
\hline
\end{tabular}
\end{center}
\hfill
\label{sfr_data}
\end{table*}
%

In Fig.~\ref{fig.sfrd_shade_all}, we also  plot the SFRD determined through different 
observations.There are different indicators of star formation but not all are independent of 
the assumed dust correction. Therefore for comparison we use the SFRDs determined through the 
radio, H-$\alpha$ and FIR emission from galaxies. We do not consider the SFRD determined with 
observations like the UV luminosity where it mainly depends on the assumed values of 
$A_{\rm FUV}$ and the extinction curve. We select the measurements where luminosity densities 
are converted to the SFRD using conversion laws (see Eq.~\ref{Eq.conversion}) with assumed 
Salpeter IMF with stellar mass with range 0.1 to 100 M$_\odot$, similar to the one we use. 
Table~\ref{sfr_data} summarizes the references to such a data and indicates the plotting 
symbols used for it in Fig.~\ref{fig.sfrd_shade_all}. Our inferred SFRD($z$) using the `cal', 
`lmc' and `mw' models are higher than these SFRD measurements with different techniques at 
$z<1$ (see, Fig.~\ref{fig.sfrd_shade_all}). The SFRD($z$) calculated for the `smc' model 
is consistent with SFRD measurements using radio observations at low $z$ but under-predict 
the SFRD at $2>z>1$. We find that the SFRD($z$) determined using the `lmc2' model fits th
e SFRD data well at all $z$. Like in the case of $A_{\rm FUV}$, determination of SFRD$(z)$ 
based on the LMC2 extinction curve provides the best fit to the different independent 
measurements compared to those of other models.  

From Fig.~\ref{fig.dust_shade_all} and Fig.~\ref{fig.sfrd_shade_all}, it is clear that, 
irrespective of the extinction law used, we find the $z$ at which the SFRD($z$) peaks is 
higher than the $z$ beyond which $A_{\rm FUV}$($z$) begins to decline. The similar trend in 
peaks of SFRD and $A_{\rm FUV}$ has reported in \citet{Cucciati12} and \citet{Burgarella13}. 

%
%
\begin{figure*}
\centering
\includegraphics[bb=80 370 545 710,width=12.0cm,keepaspectratio,clip=true]{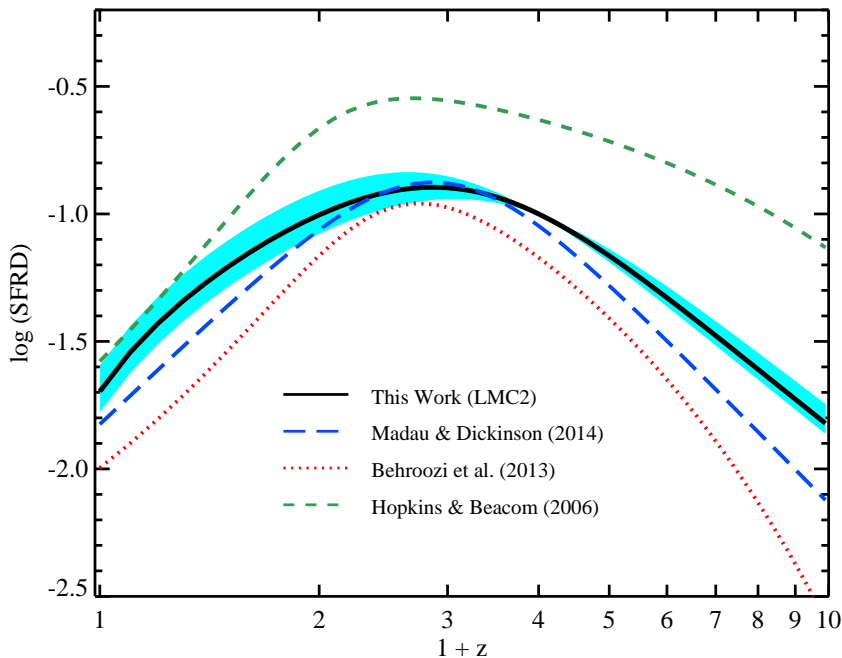}
\caption{ Our best fit SFRD($z$) in units M$_{\odot}$ 
yr$^{-1}$ Mpc$^{-3}$ obtained using the LMC2 extinction curve.  
The \emph{gray shaded} region gives the range covered by 
the low and high models. We also plot the SFRD($z$) determined 
by \citet{Madau14} 
using measurements of $\rho_{\rm FUV}$ for the same IMF 
we use but with different metallicity. 
To compare the general trend, we also plot the SFRD($z$) fit given 
by \citet{Behroozi13} for the different observational determination 
of the SFRD given in literature for the recent data and the old
data compiled by \citet{Hopkins06}. Both these fits are for 
\citet{Chabrier03} IMF. 
We scaled the \citet{Hopkins06} fit by 2.5 for clarity. }
\label{fig.sfrd_comp}
\end{figure*}
%
%
In Fig.~\ref{fig.sfrd_comp} we plot our SFRD obtained using the LMC2 extinction curve along 
with the SFRD determined by \citet{Madau14}. Shaded region in Fig.~\ref{fig.sfrd_comp} 
represents SFRD range covered when we use low and high `lmc2' models to determine the SFRD and 
$A_{\rm FUV}$. \citet{Madau14} used the $\rho_{\rm FUV}$ measurements from the literature and 
converted them in to the SFRD using the conversion constant $\zeta=1.15\times10^{-28}$ which 
is 10\% smaller than what we use. For the dust correction they use the $A_{\nu}$ provided by 
the different surveys from where the luminosity functions are used to get the 
$\rho_{\rm FUV}$. They calculate the $\rho_{\rm FUV}$ by integrating the luminosity function 
from $L_{min}=0.03L^*$, while in our case we directly take $\rho_{\rm FUV}$ given in different 
references where it is often calculated with $L_{min}=0$ (for the $L_{min}$ values used here, 
see Table~\ref{lmin_table} in appendix). \citet{Madau14} use the same IMF used by us but take 
different metallicities and consider metallicity evolution with $z$. As compared to our 
preferred SFRD($z$) for the LMC2 model the SFRD of \citet{Madau14} shows rapid increase and 
decrease in low and high $z$, respectively. However, the difference between both is within 
0.1 to 0.2 dex for $z<5$. The peak of SFRD($z$) of our preferred `lmc2' model matches exactly 
with that of \citet{Madau14}. The peak of our SFRD($z$) is at $z=1.9^{+0.2}_{-0.3}$ which is 
also consistent with the peak of SFRD reported by \citet{Cucciati12}.
%
%
\begin{table}
\caption{Fitting parameters for the SFRD($z$)$^*$}
\begin{tabular}{ l l c c c c c c}
\hline              
\hline
 Extinction      & $\rho_{FUV}$   & a & b & c & d &  \\
   curve         & fit$^\dagger$          & (10$^{-2}$) & (10$^{-2}$) &  &  &  \\
\hline
\hline
 SMC            &   Low       &   1.55 &    7.14 &    2.53 &    3.10  \\    
                &   Median    &   1.38 &    6.24 &    2.65 &    3.01  \\    
                &   High      &   1.50 &    5.12 &    3.08 &    3.09  \\  
\hline  
 LMC2           &   Low       &   2.54 &    10.9 &    2.22 &    3.07  \\    
                &   Median    &   2.01 &    8.48 &    2.50 &    3.09  \\    
                &   High      &   1.67 &    7.09 &    2.74 &    3.02  \\ 
\hline                 
 LMC            &   Low       &   3.57 &    13.6 &    2.15 &    3.13  \\    
                &   Median    &   3.13 &    9.88 &    2.37 &    3.03  \\    
                &   High      &   3.03 &    7.37 &    2.70 &    3.01  \\ 
\hline
                  
 Milky-Way      &   Low       &   6.27 &    15.2 &    2.14 &    3.16  \\    
                &   Median    &   5.78 &    11.2 &    2.28 &    3.02  \\    
                &   High      &   5.03 &    8.33 &    2.59 &    2.99  \\ 
\hline                 
 Calzetti       &   Low       &   4.44 &    15.8 &    2.06 &    3.11  \\    
                &   Median    &   3.62 &    12.0 &    2.20 &    2.97  \\    
                &   High      &   3.23 &    8.78 &    2.54 &    2.97  \\                                                                                  
\hline              
\hline
\end{tabular}
\hfill
\label{sfrd_parm}
\begin{flushleft}
\footnotesize {*The fitting form is ${\rm SFRD}(z)=\frac{a+bz}{1+(z/c)^{d}}$}  
M$_{\odot}$ yr$^{-1}$ Mpc$^{-3}$ .\\
\footnotesize {$^\dagger$Note that, as explained in the text, the models 
with low and high $\rho_{\rm FUV}$ fit 
need not to give higher and lower SFRD($z$) than median model for all $z$, respectively.}\\
\end{flushleft}
\end{table}
%

For comparing the SFRD($z$) shapes, we also plot the fit to the SFRD measurements 
compiled from the different observational data reported in the literature given by 
\citet{Behroozi13} (see figure 2 and table 4 of their paper). \citet{Behroozi13} provides 
the fit for the recent data and the old data used by \citet{Hopkins06}. Both of these fits
are obtained for \citet{Chabrier03} IMF and show rapid increase at low $z$ as compared to
our SFRD($z$). At high $z$, the fit through compiled SFRD data used by \citet{Hopkins06} 
shows the slow decrease while \citet{Behroozi13} fit shows rapid decrease as compared to our 
SFRD. The peak of the compiled SFRD measurements of \citet{Behroozi13} for the old and new 
measurements is at $z=1.7$ which is consistent with our SFRD$(z)$ peak within the allowed 
uncertainties. The main differences between our SFRD$(z)$ estimated here and the compiled 
SFRD measurements of \citet{Behroozi13} and SFRD($z$) estimated by \citet{Madau14} is that 
we self-consistently calculate the $A_{\rm FUV}(z)$ which gives SFRD$(z)$ consistent with 
$\rho_{\nu}$ measurements at different wavebands and redshifts.

Having obtained the best fit $A_{\rm FUV}$($z$) and SFRD($z$), we use the stellar population 
synthesis models to calculate the emissivity from UV to NIR regime. However, to generate the 
complete EBL, in addition to this we need the IR emissivity. We predict the IR emissivity 
using our best fit $A_{\rm FUV}$($z$) and SFRD($z$) which is explained in the following 
section.
%

\section{Galaxy emissivity in infrared}\label{sec.fir}
The old stellar population, the interstellar gas and the dust are main sources which 
contribute to the IR emission form galaxies. The IR emission from old stars peaks around 1 to 
3$\mu$m and a very few per cent of the total IR output of a galaxy is emitted by atoms and 
molecules that constitute the interstellar gas. The main source of the IR emission at 
$\lambda > 3\mu$m is a thermal emission from the dust grains heated by the local stellar 
light in the UV and optical wavelength range. The amplitude and shape of this emission from 
the NIR to FIR wavelengths depend on the temperature, size distribution and the composition 
of the dust grains \citep[e.g.][]{Dale12, Magdis12, Magdis13}. However, like the extinction 
curves, these quantities are unknown for distant galaxies. Therefore, instead of assuming the 
dust properties to model the NIR to FIR emission of a typical galaxy, we use the observed 
IR templates and make use of the $A_{\rm FUV}$ for different models determined here.

We use the average IR templates of \citet{Rieke09} from 5$\mu$m to $30$cm obtained for 
infrared galaxies having different total infrared luminosity, $L_{\rm TIR}$. They have 
assembled the SEDs of 11 local luminous and ultra-luminous infrared galaxies and for 
generating the templates at lower luminosity they have combined the templates of 
\citet{Dale07} and \citet{Smith07}. The shape of each template is moderately different for 
different $L_{\rm TIR}$. Therefore, we have to choose an appropriate template for calculating 
the IR emission. Since at any $z$, most of the total luminosity is contributed by the galaxies 
with luminosity $L^*(z)$, we choose templates of \citet{Rieke09} obtained for 
$L_{\rm TIR}=L_{\rm TIR}^*(z)$. We use $L_{\rm TIR}^*(z)$ values for different redshift using 
the total IR luminosity function given in \citet{Gruppioni13} up to $z=4$. To get the 
$L_{\rm TIR}^*(z)$ values for the high redshifts (in units $L_{\odot}$) we fit a second 
degree polynomial through $log(L_{\rm TIR}^*)$ using {\sc mpfit IDL} routine. The best fit 
second degree polynomial is $log[L_{\rm TIR}^*(z)]=10.0+1.18z-0.18z^2$. To compute the FIR 
spectrum we interpolate the templates of \citet{Rieke09} for corresponding values of 
$L_{\rm TIR}^*(z)$ given by the above polynomial fit. For all redshifts $z>7$ we use 
the interpolated template of \citet{Rieke09} with $L_{\rm TIR}^*=10^9 L_{\odot}$. However 
this lower limit has no effect on the shape of IR template since it is just a template for 
lowest luminosity ($10^{9.75} L_{\odot}$) given by \citet{Rieke09} scaled to give 
the $L_{\rm TIR}=10^9 L_{\odot}$. \footnote{Note that the range of
wavelengths used to define $L_{\rm TIR}$ is different in the case of \citet{Rieke09} (5 to 
1000 $\mu$m) and \citet{Gruppioni13} (8 to 1000$\mu$m). We take this into account and scale 
the \citet{Rieke09} templates with the definition \citet{Gruppioni13} which we use for IR 
emission from galaxies.} 

We use the energy conservation to calculate the IR emission. We assume that the average 
energy absorbed by the dust from FUV to NIR regime per Mpc$^{3}$ per $s$ at any redshift 
$z_0$, $E_{abs}(z_0)$, is emitted in the NIR to FIR regime as a thermal emission with the 
assumed SED taken from the appropriate IR template at $z_0$ as explained above. The 
$E_{abs}(z_0)$ is given by, 
\begin{equation}\label{Eq.Eabs}
E_{abs}(z_0)=\int_{\nu_i}^{\nu_f}d\nu\Big[1-C_{\nu}(z_0)\int_{z_0}^{\infty}{\rm SFRD}(z)\,\,l_{\nu}(t_{0}(z), Z)\,\,dz\,\Big],
\end{equation}
where, $C_{\nu}(z_0)=10^{-0.4\,A_{\nu}(z_{0})}$ and $\nu_i$ and $\nu_f$ are the frequencies 
corresponding to 0.092$\mu$m and 10$\mu$m. Photons in this wavelength range heat the 
interstellar dust effectively. Hard photons at $\lambda < 0.092\mu$m are mainly photo-absorbed 
by the interstellar hydrogen and helium. We assume that $E_{abs}(z_0)$ is emitted at the same 
time in IR from  5 to 1000$\mu$m. We scale the IR template with total IR luminosity 
$L_{\rm TIR}^{*}(z_0)$ to match the value of $E_{abs}(z_0)$ in between wavelength from 
5 to 1000$\mu$m. Then we do a power law extrapolation to this scaled IR template at 
$\lambda < 5\mu$m. We use a second degree polynomial to smoothly connect the NIR part 
of the SED with the IR part of the extrapolated template at the connecting points.
\footnote{ The connecting points are usually in between 2 to 4$\mu$m.} Note that, here we 
assume the efficiency of dust to re-emit in the NIR to FIR wavelengths is 100\%. 

%
%
\begin{figure*}
\centering
\includegraphics[bb=105 360 490 710,width=12 cm,keepaspectratio,clip=true]{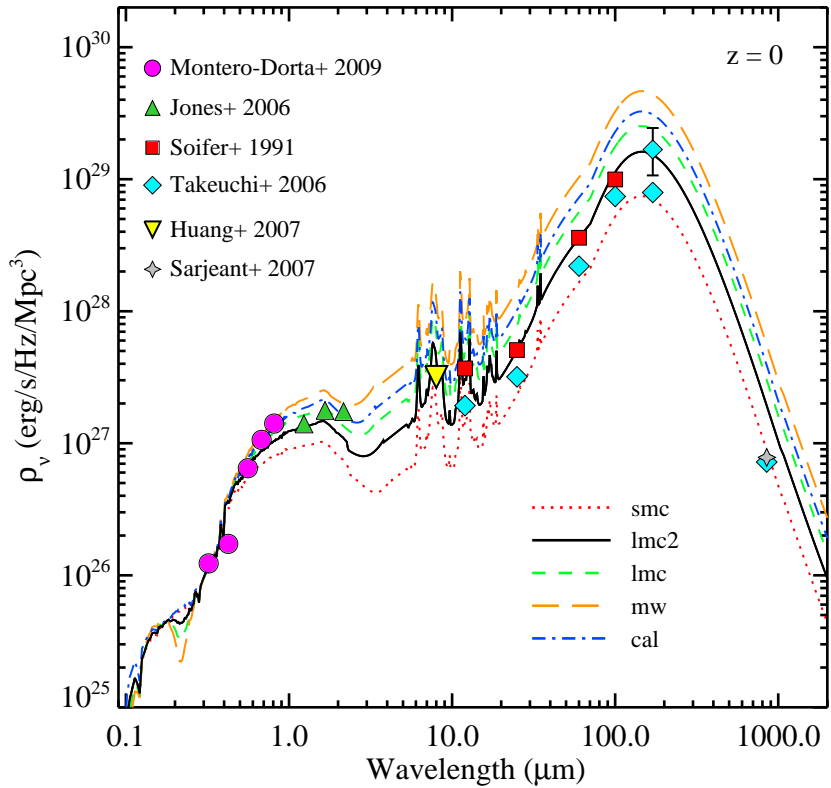}
\caption{ The galaxy emissivity at $z=0$ for different models. Apart from the 
$2175$\AA~absorption feature, in the IR ($\lambda>1\mu$m) the difference between 
different models can be easily seen. The data in the optical and NIR 
($0.35 \le \lambda < 3 \mu$m) is taken from \citet{Montero09} and  \citet{Jones06}. 
Emissivity measurements in the NIR to FIR wavelengths ($8 \le \lambda \le 850 \mu$m) 
are taken from \citet{Huang07}, \citet{Soifer91}, \citet{Takeuchi06} and \citet{Serjeant05} }
\label{fig.emis}
\end{figure*}
%
%
In Fig.~\ref{fig.emis}, we show the emissivity of galaxies from the UV to FIR
at $z=0$ for the different extinction curves assumed. As expected, apart from the 
$2175 \rm \AA$ absorption feature the emissivity up to $\lambda=0.8\mu$m is quite same for all 
models. However, from the NIR to FIR wavelengths the emissivity is different for different 
models. This is because the absorbed energy by the interstellar dust depends on the extinction
curve and the $A_{\rm FUV}$. In Fig.~\ref{fig.emis}, we also show the local emissivity from 
different surveys and at different wavelengths like in the optical from SDSS by 
\citet{Montero09}, in the NIR wavelength from the 2MASS and 6dF by \citet{Jones06}, in the IR 
to FIR from surveys like ISO FIRBACK, IRAS and SCUBA  by \citet{Soifer91}, \citet{Takeuchi06} 
and \citet{Serjeant05} and at 8$\mu$m from Spitzer space telescope survey by \citet{Huang07}. 
Our predicted local emissivity from the NIR to FIR ($\lambda > 2\mu$m) range in case of the 
`mw' and `cal' models give higher intensity by factor $\sim 2$ than the local emissivity 
measurements from these different surveys. Our local emissivity for the `lmc' and `smc' models 
are marginally higher and lower from these measurements, respectively. Our `lmc2' model gives 
a local emissivity which is in very good agreement with these measurements. 

In principle, the dust can be assumed to have lower efficiency to re-emit. In that case, the 
models which give higher IR emissivity can be scaled down to reproduce the measurements. But 
there is no room for scaling up the IR emission form the models which give lower IR 
emissivity. Therefore, the `smc' model can not be scaled up to match the local emissivity 
measurements.

We use the emissivity of galaxies from UV to FIR range obtained here to calculate the EBL 
which is discussed in detail in the following section.
%
%
\section{EBL Calculation}\label{sec.eblcal}

We solve the cosmological radiative transfer equation (Eq.~\ref{Eq.rad_t}) numerically to 
compute the EBL. In this equation the source term, $\epsilon_{\nu}$, is the sum of the QSO 
and galaxy emissivities. We take the QSO emissivity, $\epsilon_{\nu , \rm Q}(z)$, and the 
SED as given in Section~\ref{sec.ebl}. We use the galaxy emissivity from the combinations of 
the SFRD($z$) and A$_{\rm FUV}(z)$ for different dust extinction curves and corresponding 
dust emission as described in the previous sections. In the following sections, we compare our 
estimated EBL from UV to FIR regime with the direct measurements of the local EBL and other 
estimates of the EBL reported in the literature.   
\subsection{The local EBL}\label{sec.firebl}
\begin{figure*}
\centering
\includegraphics[bb=95 360 550 711,width=12 cm,keepaspectratio,clip=true]{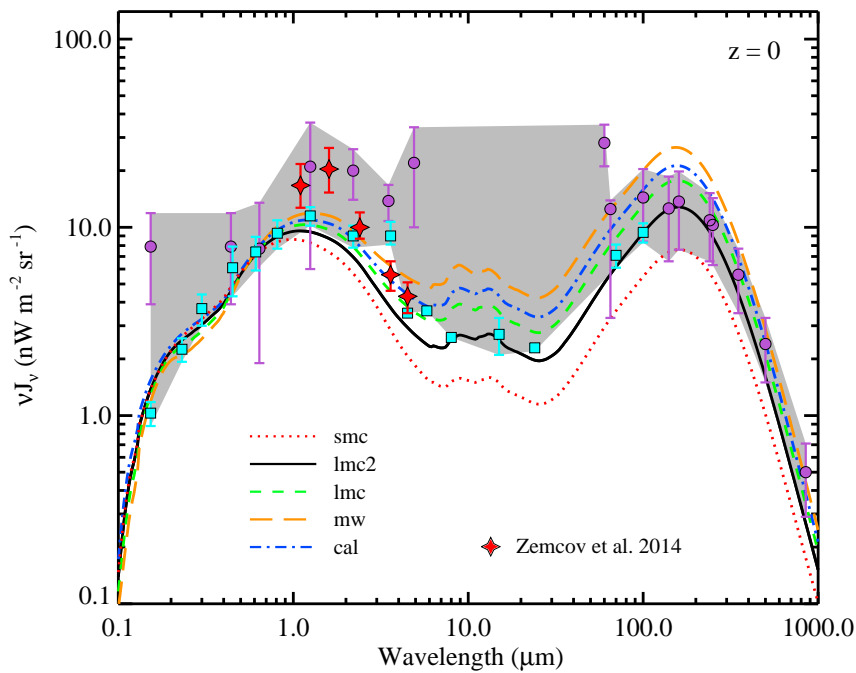}
\caption{ The EBL at $z=0$ for different models. 
The shaded region represents the range of the allowed EBL intensity from observations. 
The lower limits are determined by the intensity of the IGL (\emph{squares}). 
While the upper limits are determined by the direct measurements of the EBL (\emph {circles}). 
The data used here is taken from \citet{Dwek13} and references their in. The IR 
predictions of the `smc'
model are inconsistent with the observations. The \emph{stars} 
are the sum of the IGL background
compiled by \citet{Franceschini08} and the estimated intra-halo light 
from the measured fluctuations of EBL by
\citet{Zemcov14}.}
\label{fig.ebl}
\end{figure*}

First, we compute the EBL at $z=0$ for our five different models which are plotted in 
Fig.~\ref{fig.ebl}. Apart from the wavelengths near the 2175\AA~absorption, different EBLs 
are quite indistinguishable from each other for $\lambda < 0.8\mu$m. As expected, we see a 
clear differences appearing at higher wavelengths between different models. Apart from the 
extinction curves, the difference is also because of the differences in $A_{\rm FUV}$ values 
for the different models (see Fig.~\ref{fig.dust_shade_all}). Therefore, even though the 
estimated EBL in the UV and optical parts are similar (because of the way we determine SFRD 
and $A_{\rm FUV}$), there are clear differences in the IR wavelengths where the dust 
re-emission is important.

%
%
The shaded area in Fig.~\ref{fig.ebl} represents the range of the allowed EBL intensity 
determined by the local EBL observations. The data used in Fig.~\ref{fig.ebl} is taken from 
compilation of \citet{Dwek13} (see their Fig. 7). The lower limits are determined by the 
intensity of integrated galaxy light (IGL). The IGL is obtained by adding the light emitted 
by resolved galaxies in deep surveys. In principle the IGL should converge to the total EBL 
at $z=0$. However, because of the problem in the convergence of number counts and sensitivity 
of the surveys to resolve fainter galaxies, the IGL gives a lower limit to the EBL. Here, we 
use the IGL measurements in the UV from Galex (Galaxy Evolution Explorer), in the optical 
from HST and in the IR from Spitzer, ISO and Herschel \citep{Totani01, Xu05, Levenson07, 
Fazio04, Bethermin10, Berta10}. The upper limits on the local EBL come from the direct 
measurements of the EBL. Important uncertainty in the direct measurements is the removal of 
strong foreground zodiacal light caused by the interplanetary dust and the stellar emission 
from the Milky-way. Therefore, these measurements provide a strict upper limits on the local 
EBL. In Fig.~\ref{fig.ebl}, we take the absolute measurements of EBL in the optical from 
Pioneer 10/11 and in the IR from COBE 
\citep{Fixsen98, Dwek98, Finkbeiner00, Dole06, Levenson07, Matsuoka11, Matsuura11}.

Our EBL estimate for the `smc' model goes below the lower limits at $\lambda>1\mu$m. While 
for all other models, the estimated EBL is within the allowed range. If we strictly follow 
the shaded region, we can rule out the `smc' model and conclude that the average extinction 
curve for galaxies is inconsistent with the SMC type of dust extinction. However, the allowed 
range is too large to distinguish between other models. All our EBL models in the UV, optical 
and NIR ($\lambda <2 \mu$m) follow the lower limits of the EBL where we use the emissivity 
consistent with the observed multi-wavelength galaxy luminosity functions. The `lmc2' model, 
which also provides the $A_{\rm FUV}(z)$ and SFRD$(z)$ consistent with different independent
measurements, produces the IR background ($1<\lambda<100 \mu$m) consistent with the lower 
limits while other models produce slightly higher background intensity but well within the 
allowed range and closer to the lower limits. In the FIR regime ($\lambda >100 \mu$m), the 
estimated background for the `lmc2' and `lmc' model goes through the observed points. The 
EBL for the `mw' and `cal' models are just consistent or slightly higher than the observed 
upper limits at $\lambda >100 \mu$m. In summary, the available local EBL measurements in
the NIR to FIR regime doest not support the average dust extinction similar to the one 
observed in case of SMC. However, these measurements can not discriminate between the EBL
obtained with other extinction curves.

Recently, using the rocket-borne instrument Cosmic Infrared Background Experiment 
(CIBER), \citet{Zemcov14} have measured the fluctuation amplitude of IR background at 
$1.1\mu$m and $3.6\mu$m. One of the plausible explanations for the large fluctuation found 
at these wavelengths is that it arises from the intra-halo light (IHL) produced by the 
tidally striped old stars in the halo of the galaxy \citep{Cooray04, Thacker14}. Using these 
fluctuations measured over the large scales \citet{Zemcov14} gives the model dependent 
total EBL contributed by IHL. In Fig.~\ref{fig.ebl}, we show their computed values of the 
total EBL which is sum of their estimated IHL and the compiled measurements of IGL by 
\citet{Franceschini08}. As expected, since we do not include the additional contributions 
like the IHL in our emissivities, we find that our estimated EBL for all models is lower 
than the predicted by \citet{Zemcov14}. Except for the `smc' model, the EBL obtained for all 
other models at $\lambda \le 2.4\mu$m are within 2-$\sigma$ lower than the total EBL 
predicted by \citet{Zemcov14}. We find that only for the `smc' model it is more than 
2.5-$\sigma$ lower at all wavelengths. At $\lambda \ge 3.6\mu$m our `lmc', `mw' and 
`cal' models match with the EBL predicted by \citet{Zemcov14} within 1-$\sigma$.
In the light of these recent developments, it will be interesting to consider the IHL 
contribution to IR, the signal arising from the epoch of re-ionization 
\citep{Cooray04, Kashlinsky04} and the FIR light from dusty galaxies 
\citep{Amblard10, Thacker13, Viero13} which we will attempt in the near future.

From the very good agreement with the local emissivity (see Fig.~\ref{fig.emis}) and with 
different independent measurements of the $A_{\rm FUV}(z)$ and SFRD$(z)$ (see 
Fig.~\ref{fig.dust_shade_all} and \ref{fig.sfrd_shade_all}), for the EBL calculations we 
prefer our `lmc2' model over other models.   

\subsection{High $z$ EBL}\label{sec.firebl}
\begin{figure*}
\centering
   \includegraphics[bb=90 390 560 710, width=17.5cm, keepaspectratio, clip=true]{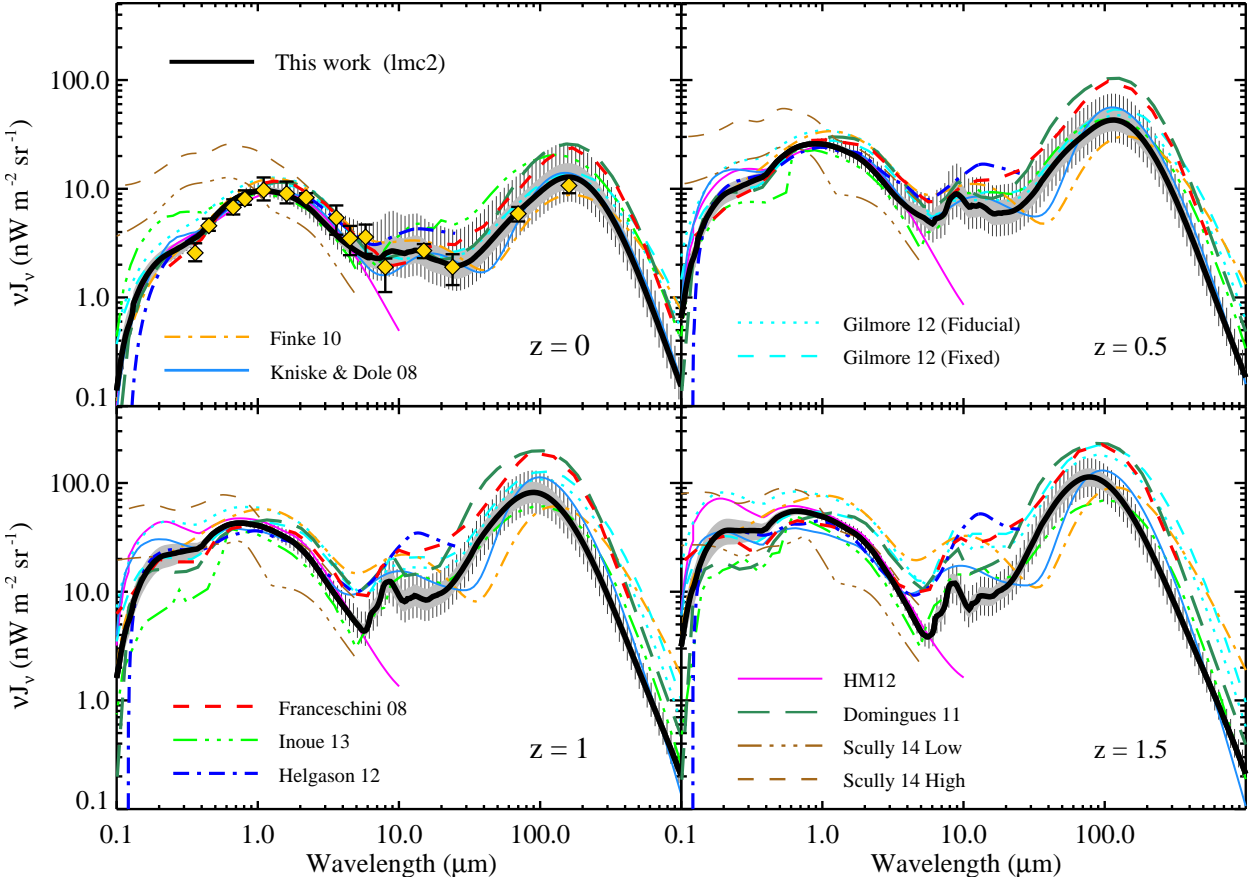}
  \caption{ The predicted EBL at different redshifts for our preferred 
`lmc2' model. The \emph{gray shaded} region shows
the EBL covered by the high and low `lmc2' model. The \emph{vertical striped} 
region gives the range covered 
by all other median models with different extinction curves.
For comparison we plot other estimates of the EBL from literature. 
In the top left panel the \emph{yellow diamonds} are the lower 
limit data compiled by \citet{Kneiske10}.
For clarity the legends are distributed over the entire plot. 
Note that the EBLs presented here are in proper units. }
\label{fig.eblall}
\end{figure*}

In Fig.~\ref{fig.eblall}, we plot the EBL at redshifts 0, 0.5, 1 and 1.5  for our preferred 
`lmc2' model. We also show the range covered by the high and low `lmc2' model by a gray 
shaded region and the range covered by all five median models by a vertical striped region. 
The fact that we made sure the SFRD($z$) and A$_{\rm FUV}(z)$ determined for different 
extinction curves should give the same observed emissivity has resulted in the narrow spread 
in the striped region for $\lambda <3\mu$m, especially at high redshifts. For comparison, in 
Fig.~\ref{fig.eblall}, we also show the previous estimates of the EBL reported in the 
literature \citep[][HM12]{Inoue13, Gilmore12, Finke10, Kneiske10, Franceschini08,  
Dominguez11, Helgason12, Scully14}. Below we compare our EBL predictions with the previous 
EBL estimates which use observational data like galaxy number counts and galaxy luminosity 
functions to get the EBL directly.
 
%
%
%
%
\citet{Helgason12} and \citep{Stecker12} reconstructed the EBL using the multi-wavelength and 
multi-epoch luminosity functions. We also use a similar compilation of luminosity functions 
but up to the $K$ band. Therefore, our EBL matches very well with the EBL predictions of 
\citet{Helgason12} up to $\lambda \sim 3 \mu$m, as shown in Fig.~\ref{fig.eblall}. For 
$\lambda > 3 \mu$m, \citet{Helgason12} predicts the higher EBL intensity than us.
The EBL model of \citep{Scully14} extend the model of \citep{Stecker12} upto 5$\mu$m and 
provide 1-$\sigma$ upper and lower limits on the EBL as well as $\tau_{\gamma}$. Our 
EBL predictions are consistent with their lower and upper limit EBL for $\lambda >1\mu$m. 
The EBL estimated by \citet{Dominguez11} is based on the observed K-band luminosity function 
with the galaxy SEDs based on the multi-wavelength observations from the SWIRE library. 
For $z<1$ and $\lambda \sim 3 \mu$m, our EBL is consistent with the predictions of 
\citet{Dominguez11} which shows slightly lower EBL intensities in the UV and slightly higher 
EBL intensities in the NIR. At high $z$ this difference is more prominent.
\citet{Franceschini08} used different multi-wavelength survey data which includes 
luminosity functions, number counts and the redshift distribution of different galaxy types 
and the relevant data is fitted and interpolated to get the EBL. Our estimated EBL matches 
well with the  EBL of \citet{Franceschini08} up to NIR wavelengths. At $z>1$, in the UV 
regime their EBL gives factor $\sim 1.5$ lower intensity than our EBL. However in the FIR 
wavelengths, our EBL gives around factor $\sim ~2$ smaller EBL intensity as compared to the 
EBL of \citet{Franceschini08} and \citet{Dominguez11}.
In summary, as expected our EBL at $\lambda <3 \mu$m is consistent with the models which use 
direct observations to get the EBL but gives a lower EBL in the NIR to FIR regime. 
Below we mention some of the general trends in different EBL estimates that can be seen 
from  Fig.~\ref{fig.eblall}.

Given that there are more observations of the EBL as well as the luminosity functions of 
galaxies in the local Universe, almost all the different independent models including our 
`lmc2' model converge very well, spans narrower range at $z=0$ in the UV to NIR regime and 
pass through observed lower limits of EBL. However, at the NIR and FIR wavelengths the spread 
between different estimates is relatively higher. All the EBL estimates differ from each 
other at high $z$ and the difference is as high as factor $\sim$4. Our local EBL passes very well 
through the lower limit EBL data compiled by \citet{Kneiske10}. Most of the EBL models for 
$0.4 <\lambda < 2 \mu$m give similar intensity upto $z<1.5$. 

Our local EBL is very much similar to the estimates of HM12. However, at higher $z$, the UV 
background ($\lambda <0.4 \mu$m) intensity of HM12 is higher than our EBL. This will have 
implications on the values of escape fraction for H~{\sc i} ionizing photons which indirectly 
plays an important role in interpreting He~{\sc ii} Lyman-$\alpha$ effective optical depth 
measurements near the epoch of Helium reionization \citep[see,][]{KS13}.
 
Having obtained the EBL at different $z$ from UV to FIR, we calculate its effect on the 
transmission of high energy $\gamma$-rays through the IGM and compare it with the different 
observations in the following section. 
\section{Gamma ray attenuation}\label{sec.gamma-tau}

Two photons with sufficient energy upon collision can annihilate into an electron positron 
pair. The condition on energies of photons ($E_1$ and $E_2$) for this process of pair 
production is given by 
\begin{equation}
\sqrt{2E_1 E_2 (1-\cos\theta)} \geq 2 m_e c^2,
\end{equation}
where, $\theta$ is the collision angle, $m_e$ is the mass of the electron and $c$ is the 
speed of light. Thus $\gamma$-rays with an energy $E_\gamma$ can annihilate themselves with 
the background extra-galactic photons having energy greater than a threshold energy $E_{th}$,
\begin{equation}
E_{th}=\frac{2m_e^2c^4}{E_\gamma (1-\cos\theta)}.
\end{equation}
The cross-section for this process is,
\begin{eqnarray}
\lefteqn{\sigma(E_1,E_2,\theta) = \frac{3\sigma_T}{16}(1-\beta^2)} \nonumber \\
& & \times \left[ 2\beta(\beta^2-2)+ (3-\beta^4)\ln \left( \frac{1+\beta}{1-\beta}\right)\right], 
\label{Eq.cross} 
\end{eqnarray}
where,
\[\beta = \sqrt{1-\frac{2m_e^2c^4}{E_1 E_2 (1-\cos\theta)}}\,\,,\]
and $\sigma_T$ is the Thompson scattering cross-section. The pair production cross-section 
given in Eq.~\ref{Eq.cross} has a maximum value $\sigma(E_1,E_2,\theta)_{max}=0.25\,\sigma_T$ 
and the corresponding value of $\beta=0.7$.

If the number density of background photons at redshift $z$ and energy 
$E_{bg}$ is $n(E_{bg},z)$ (from Eq.~\ref{Eq.num}), then as a result of pair production the 
optical depth encountered by the $\gamma$-ray photons emitted at redshift $z_0$ and observed 
at energy $E_\gamma$ on the Earth (i.e. at $z=0$) is given by
\begin{eqnarray}
\lefteqn{\tau_{\gamma}(E_\gamma,z_0) =  \frac{1}{2}\int^{z_0}_0 dz\;\frac{dl}{dz}\int^1_{-1}d(\cos\theta) \; (1-\cos\theta)} \nonumber \\ 
& & \times \int^{\infty}_{E_{min}} dE_{bg}\; n(E_{bg},z)\;\sigma(E_\gamma (1+z),E_{bg},\theta).
\label{Eq.tau}
\end{eqnarray}
Here,  
\begin{equation}
E_{min}=E_{th}\:(1+z)^{-1}=\frac{2m_e^2c^4}{E_\gamma (1+z)(1-\cos\theta)}\,\, .
\end{equation}
Above equation in terms of the maximum wavelength of the EBL, which is going to attenuate 
observed $\gamma$-rays of energy $E_{\gamma}$, can be simplified as 
$\lambda_{max}(z)=23.74 \text{\AA} E_{\gamma} (1+z)(1-\cos\theta)$ where
$E_{\gamma}$ is in GeV.  The cross-section for the pair production will be maximum at
$\lambda(z)=12.06\text{\AA} E_{\gamma} (1+z) (1-\cos\theta)$.  

The specific number density of the EBL photons is directly related to the optical depth 
$\tau_{\gamma}$ encountered by $\gamma$-rays while traveling through the IGM as explained 
above. For the EBL estimated here, we calculate $\tau_{\gamma}$ using Eq.~\ref{Eq.tau} 
over a energy range from GeV to TeV. In the following sub-section, we compare our calculated 
of $\tau_{\gamma}$ with those obtained using different EBL estimates reported in the 
literature. 
%
%
\begin{figure*}
\centering
   \includegraphics[bb=52 375 650 710  width=15cm, clip=true]{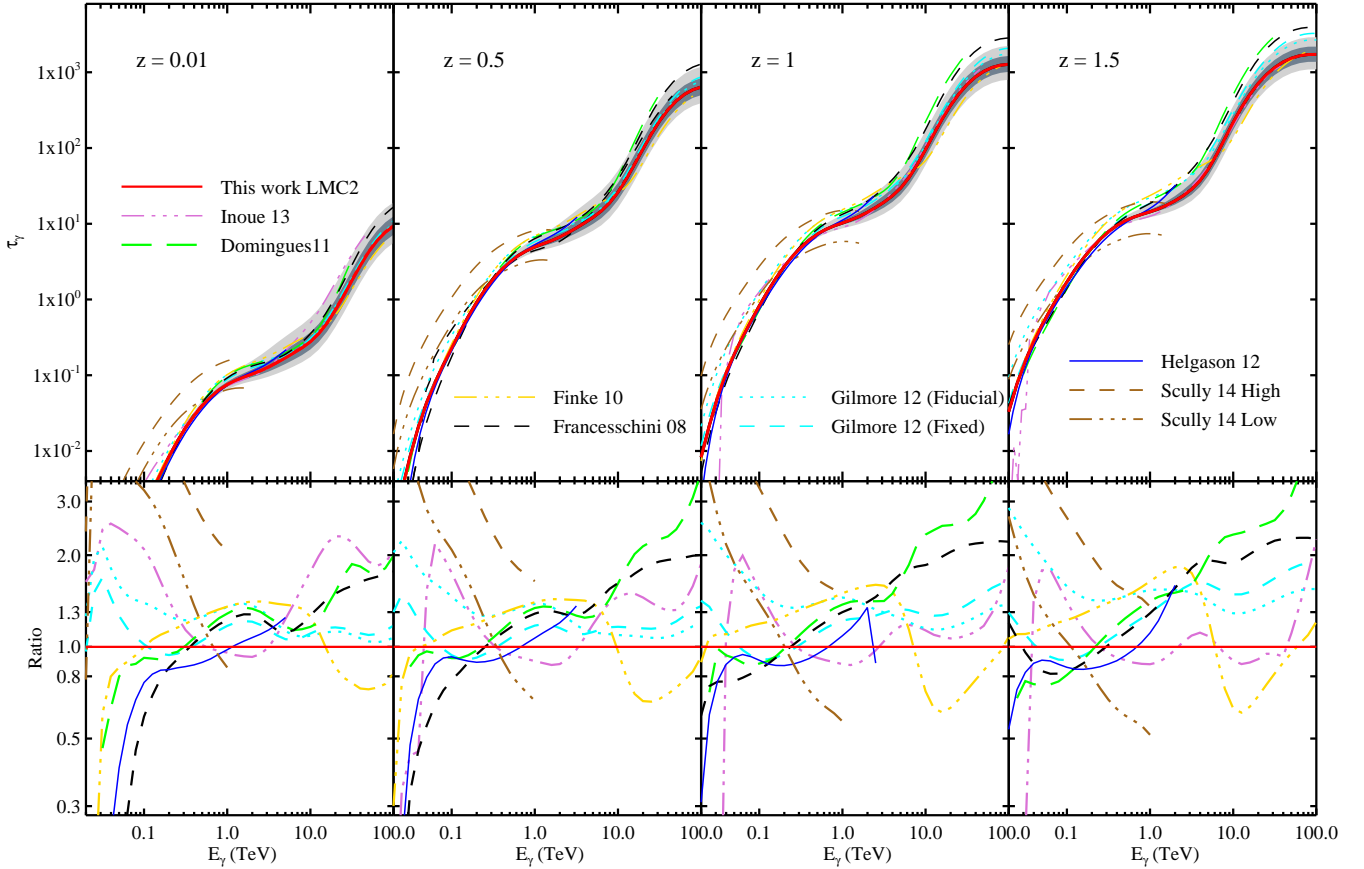}
  \caption{ \emph{Top panel}: The $\gamma$-ray optical depth, $\tau_{\gamma}$, for different 
emission redshifts for observed $\gamma$-ray energy using our preferred `lmc2' EBL 
(\emph{red curve}) along with the $\tau_{\gamma}$ from other EBL estimates from the 
literature. For clarity the legends are distributed over the entire plot. Dark gray shaded 
region gives the range covered in $\tau_{\gamma}$ by the high and low `lmc2' EBL models. The 
light gray shaded region gives the range covered by all five of our EBL models.
\emph{Bottom panel}: The ratio of $\tau_{\gamma}$ predictions of different models to the 
$\tau_{\gamma}$ obtained for our `lmc2' model.}
\label{fig.tau_z}
\end{figure*}
%
\subsection{The $\gamma$-ray opacity at different $z$ }\label{sec.taucomp}
The optical depth encountered by $\gamma$-rays while traveling from the emission redshifts 
$z_0$ to earth (i.e, $z=0$) and observed at $\gamma$-ray energies from GeV to TeV range is 
plotted in Fig.~\ref{fig.tau_z} for different emission redshifts. We plot $\tau_{\gamma}$ 
calculated for our median `lmc2' model along with the $\tau_{\gamma}$ for its low and high 
counterparts in a \emph{dark gray} shaded region. We also show the range of $\tau_{\gamma}$ 
covered by all five median models by a \emph{light gray} shaded region. The extent of the 
\emph{light gray} shaded region for different $z$ up to the $\gamma$-ray energy of 0.6 TeV 
shows that the $\tau_{\gamma}$ is indistinguishable for all five EBL models. This is because 
the maximum effective wavelength of EBL photons which interact with $\gamma$-rays of energy 
less than 0.6 TeV is less than $3\,\,\mu$m  where, by construct, our different EBL models 
predict similar intensities (see Fig.~\ref{fig.eblall}). The spread of shaded region for 
$\gamma$-ray energy $>$ 0.6 TeV points to the fact that our EBL intensity at FIR wavelengths 
is different for different models. 

For comparison, in Fig.~\ref{fig.tau_z}, we also plot the $\tau_{\gamma}$ obtained by other 
estimates of EBL given in the literature. To clearly show the differences between 
$\tau_{\gamma}$ calculated for other EBL estimates and for our `lmc2' EBL model, we plot the 
ratio of the former to the later in the bottom panel of Fig.~\ref{fig.tau_z}. Differences in 
the $\tau_{\gamma}$ obtained for various EBL estimates can be directly understood from the 
differences in the EBLs as shown in  Fig.~\ref{fig.eblall}. Here, we mention some of the 
general trends seen in the various $\tau_{\gamma}$ estimates and then compare our 
$\tau_{\gamma}$ with the models which used direct observations of galaxy properties to 
construct the EBL.

The difference between the $\tau_{\gamma}$ at energies from GeV to TeV for various EBL 
estimates increases with $z$. The difference is more in the TeV energy range where the EBL 
photons which effectively attenuate the $\gamma$-rays are from the FIR part of the EBL. 
The large scatter in different EBL estimates in the FIR wavelengths (see, 
Fig.~\ref{fig.eblall}) is responsible for that. However, for the observationally relevant 
$\gamma$-rays which are the ones with $0.1<\tau_{\gamma}<2$ where in the corresponding energy 
range the differences between $\tau_{\gamma}$ estimates are quite small. The difference in 
$\tau_{\gamma}$ for $\gamma$-ray energies from 0.1 to 3 TeV is small and within 30\% 
for various estimates for emission redshifts $z<1.5$. In general, our EBL gives lower 
$\tau_{\gamma}$ compared to most of the other estimates. The $\tau_{\gamma}$ obtained for 
lower limit EBL of \citet{Scully14} at $z=0.01$ is factor $\sim 2$ higher at 
$E_{\gamma}<0.2$TeV. The $\tau_{\gamma}$ obtained for the EBL by \citet{Inoue13} at $z\ge 1$ 
is within the 10\% of our estimate at $0.2<E_{\gamma}<40$TeV. However, note that the 
$\tau_{\gamma}$ at $E_{\gamma}>10$ TeV in our model is quite small because we do not consider 
the CMBR in our EBL model \citep[see][]{Stecker06}. The $\tau_{\gamma}$ in the energy range 
$0.05<E_{\gamma}<2$~TeV estimated by \citet{Helgason12} upto $25\mu$m is within 15\% of our 
$\tau_{\gamma}$. The $\tau_{\gamma}$ calculated using the EBL generated by 
\citet{Franceschini08} and \citet{Dominguez11} for $0.1<E_{\gamma}<4$~TeV is within the 
30\% of our $\tau_{\gamma}$. However, at high energies, $\tau_{\gamma}$ differs more than 
this which is evident from the differences in the FIR part of the EBL in between their 
models and our `lmc2' model (see, Fig.~\ref{fig.eblall}). 

%
\begin{figure*}
\centering
    \includegraphics[bb=93 360 500 711,width=11 cm,keepaspectratio,clip=true]{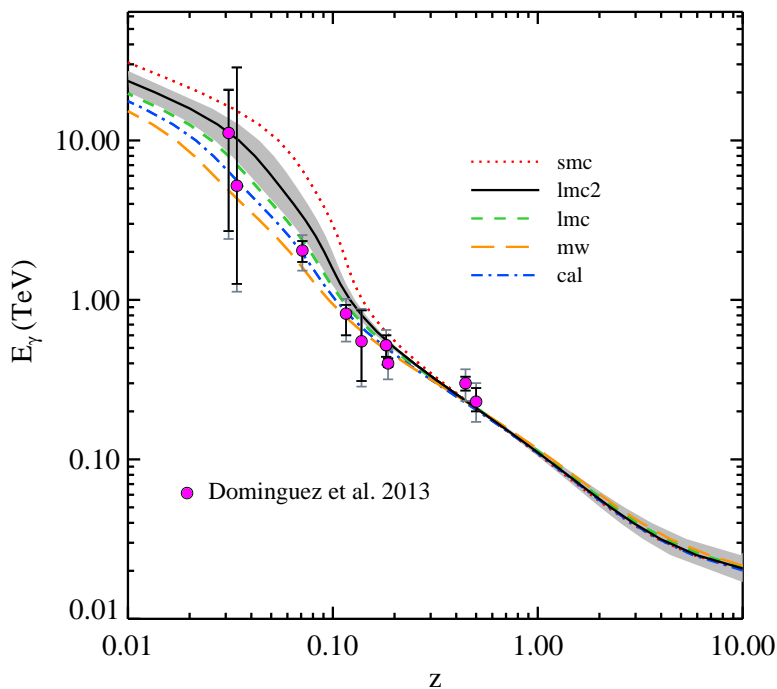}
\caption{ The $\gamma$-ray horizon plotted for different EBL models. Data points 
are from \citet{Dominguez13}. 
\emph{Gray area} shows range covered by EBL in the high and low `lmc2' models.}
\label{fig.horizon}
\end{figure*}
%
\subsection{ $\gamma$-ray horizon }\label{sec.horizon}
The $\gamma$-ray horizon is defined as the maximum redshift of the $\gamma$-ray source that 
can be detected through $\gamma$-rays of observed energy $E_{\gamma}$ with 
$\tau_{\gamma} \le 1$. This is nothing but the $\gamma$-ray source redshift $z_0$ for 
$\gamma$-rays of observed energy $E_{\gamma}$ on earth corresponding to
$\tau_{\gamma}(E_{\gamma}, z_0)=1$. In Fig.~\ref{fig.horizon}, we plot the $\gamma$-ray 
horizon for different source redshifts for our different EBL models where 
the gray shaded area shows the range spanned by the high and low `lmc2' model.

It can be directly seen from the Fig.~\ref{fig.horizon} that our universe is transparent 
for the $\gamma$-rays with energies less than 10 GeV. The $\gamma$-rays with energy 
$E_{\gamma}<$ 30 GeV can be observed from  sources at $z\sim 3$ without significant 
attenuation. Due to the differences in the IR part of the EBL, the $\gamma$-ray horizon 
redshift for our models differ in TeV energies. The well measured $\tau_{\gamma}$ at TeV 
energies at low $z$ can, in principle, distinguish between different EBL models 
presented here. 

In Fig.~\ref{fig.horizon}, we also plot the measurements of $\gamma$-ray horizon by 
\citet{Dominguez13} where they model the multi-wavelength observations of blazars to determine 
$\gamma$-ray horizon which is EBL independent. Apart from the `smc' model, all our other 
models show good agreement with these $\gamma$-ray horizon measurements. This is evident from 
the $\chi^2$ statistics. The reduced $\chi^2$ values are 13.1, 1.7, 0.5, 0.4 and 0.4 for our 
`smc', `lmc2', `lmc', `mw' and `cal' models, respectively. This indicates that for the `smc' 
model, the predicted NIR to FIR part of the EBL intensity is less and inconsistent with the 
available $\gamma$-ray horizon measurements. 

Note that the reduced $\chi^2$ quoted above includes only observational and systematic errors. 
We notice that even when we allow for the EBL obtained using high and low models for `smc', it 
does not give $\gamma$-ray horizon consistent with the measurements. If we include the average 
deviation in the $\gamma$-ray horizon in case of `lmc2' model due to its higher and lower 
limits of the EBL obtained by using high and low models as the errors in the prediction of 
$\gamma$-ray horizon, the reduced $\chi^2$ for `lmc2' becomes 0.92. It is evident 
from  Fig~\ref{fig.horizon} that, even though we could rule out the `smc' model, 
purely based on the available $\gamma$-ray horizon measurements alone, we can not 
distinguish between other four models. Given the uncertainties involved in the modeling  
intrinsic SEDs of $\gamma$-ray sources, the contribution of IHL to the galaxy emissivity 
in the NIR  and the small spread of $\gamma$-ray horizon predicted in our remaining four 
models, it may be challenging to distinguish them based on more of such measurements.

%
\subsection{ Fermi measurements of $\tau_{\gamma}$ }\label{sec.fermi}
%
\begin{figure}
\centering
    \includegraphics[bb=160 360 500 730, width=11cm, clip=true]{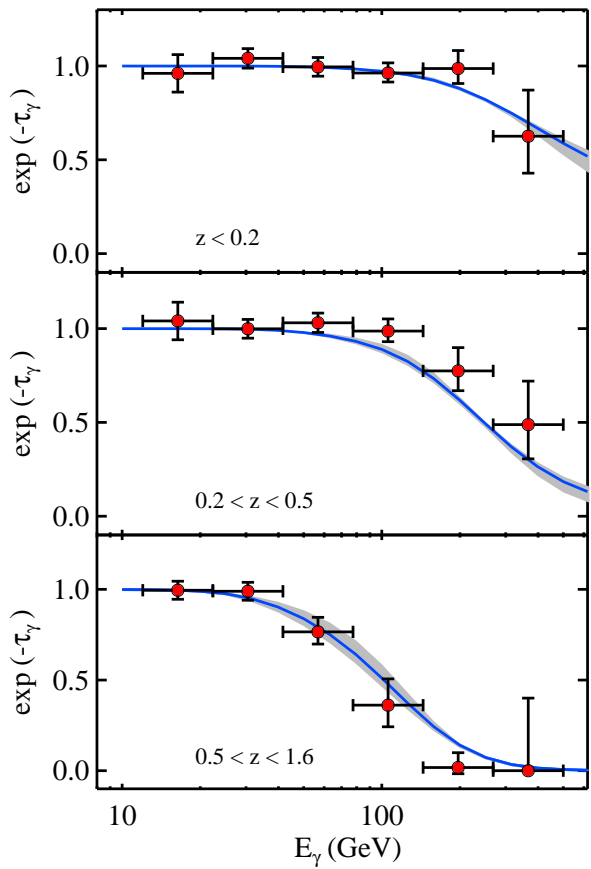}
\caption{The $\gamma$-ray transmission ($e^{-\tau_{\gamma}}$) for our 
`lmc2' EBL model (\emph{solid blue curve}) 
along with the measurements of \citet{GammaSci} in the redshift 
bins defined by them. We calculate $e^{-\tau_{\gamma}}$ for our EBL models 
which mimic the method of calculating it by stacking the individual 
blazar spectrum in a redshift bin as done by 
\citet{GammaSci} (see text in section~\ref{sec.fermi}). The 
range covered by all five EBL models along with their high and 
low counterparts is shown by \emph{gray shade}.} 
\label{fig.sci}
\end{figure}
%
Recently, \citet{GammaSci} reported the average measurements of $\tau_{\gamma}$ over a 
large redshift range using a sample of 150 blazars observed with the Fermi-LAT. Since, it is 
difficult to determine $\tau_{\gamma}$ in individual blazar spectrum, they divided their 
blazar sample into three redshift bins $z<0.2$, $0.2<z<0.5$ and $0.5<z<1.6$. Then they stacked 
spectra in each redshift bin and determined the intrinsic SED of blazars by extrapolating the 
stacked spectrum from the lower energies where various EBL estimates suggests that 
$\tau_{\gamma}$ is negligible. Using these stacked spectra, \citet{GammaSci} reported the 
measurements of $\tau_{\gamma}$ for the observed $\gamma$-ray energies from 10 to 500 GeV in 
the redshift range of $0 < z \le 1.6$. Based on the good agreement between the EBL measured 
by Fermi and that expected from the lower limits determined from the IGL measurements, they 
argued that there is a negligible room for residual emission from other sources. The 
$\gamma$-rays in this observed energy range from 10 to 500 GeV will be attenuated effectively 
by the EBL photons of rest wavelength $\lambda< 3.1\mu$m for $z<1.6$, 
$\lambda< 1.8\mu$m for $z<0.5$ and $\lambda< 1.4\mu$m for $z<0.2$ where the cross-section of 
pair-production is maximum (at $\theta=\pi$). This is the wavelength range where, by 
construct, our all five models give similar EBL.   

For comparison with the measurements of \citet{GammaSci}, we calculate the $\tau_{\gamma}$ 
in a way that mimics stacking of individual blazar spectra as done by them. We take the 
number distribution of blazars as a function of redshift for 150  blazars used by 
\citet{GammaSci} and calculate $e^{-\tau_{\gamma}}$ for each blazar at a corresponding 
$z$ and at different energies $E_{\gamma}$. Then we take the average of  
$e^{-\tau_{\gamma}}$ over the same number of blazars in the redshift bins. This average is 
equivalent of getting  $e^{-\tau_{\gamma}}$ by stacking the individual blazar spectrum in a 
redshift bin. In Fig.~\ref{fig.sci}, we plot our estimates of $\gamma$-ray optical depth 
along with the measurements of \citet{GammaSci}, in terms of the transmission 
$e^{-\tau_{\gamma}}$ for our `lmc2' EBL model. The range in $e^{-\tau_{\gamma}}$ covered by 
all five EBL models with their high and low counterparts is shown by gray shaded region 
in Fig.~\ref{fig.sci}. For our all EBL models, the $e^{-\tau_{\gamma}}$ fits well in low 
($z \le $0.2) and high (0.5$< z \le$1.6) redshift bins. However, in the intermediate 
redshifts, our estimated $\tau_{\gamma}$ is slightly higher than that reported in 
\citet{GammaSci}. This excess optical depth is not statistically significant given the large 
uncertainties in the measurements of $\tau_{\gamma}$. However if these measurements are indeed 
very accurate then to account for this we need the EBL intensity to be factor 2 lower than the 
predicted by our models at $z<0.5$ in optical to NIR regime. This requires a factor 2 
reduction in $\rho_{\nu}$ at $\lambda< 1.8\mu$m for $z<0.5$. One can, in principle, reduce the 
$\rho_{\nu}$ by increasing $L_{min}$ (in Eq.~\ref{rho}). However, how much $\rho_{\nu}$ can 
be reduced depends upon the $\alpha$ and the luminosity of the faintest galaxy detected to 
determine the luminosity functions. It can be seen from Table~\ref{lmin_table} that the 
values of $\alpha$ at $z<0.5$ are high and therefore to reduce $\rho_{\nu}$ by factor 2 one 
needs to take $L_{min} >0.2L^*$. However, in this wavelength range (UV to NIR), given the fact 
that our EBL models are consistent with the observational lower limits on local EBL, there is 
not much room available to reduce the EBL intensity. This re-iterates the findings of 
\citet{GammaSci} that the observed luminosity densities of galaxies are just sufficient to 
reproduce the $\tau_{\gamma}$ measurements. 

Note that to obtain intrinsic blazar spectra, the continuum extrapolation from the lower 
energy to higher energy is a practical solution but may not be the valid one. Therefore, the 
minor discrepancy mentioned above does not conclude anything significantly. However, it will 
be more interesting for constraining various EBL models if the errors on $\tau_{\gamma}$ are 
reduced significantly. 
      
%
\section{Effect of uncertainties on model parameters}\label{sec.res}
In this paper, we use a progressive fitting method to determine the combinations of 
$A_{\rm FUV}$($z$) and SFRD($z$) for five different extinction curves using multi-wavelength 
multi-epoch luminosity functions. We use these $A_{\rm FUV}$($z$) and SFRD($z$) to get the IR 
emissivity and generate the EBL. The progressive fitting method uses the convolution integral 
(see Eq.~\ref{Eq.convolution}) to get the $\rho_{\nu}$. The convolution integral involves the 
stellar emission from the population synthesis model which depends on the assumed input 
parameters like metallicity, IMF and age of the galaxy. In this section, we investigate the 
possible uncertainties arising from the scatter in these input parameters involved in the 
modeling as compared to the that arising purely from uncertainties in the $\rho_{\nu}$ 
measurements. In particular, we concentrate on the assumed $z_{max}$ which corresponds to 
the age of galaxies in convolution integral and the metallicity. 
%
\begin{figure*}
\centering
  \includegraphics[bb=65 370 545 710,width=12.6cm,keepaspectratio,clip=true]{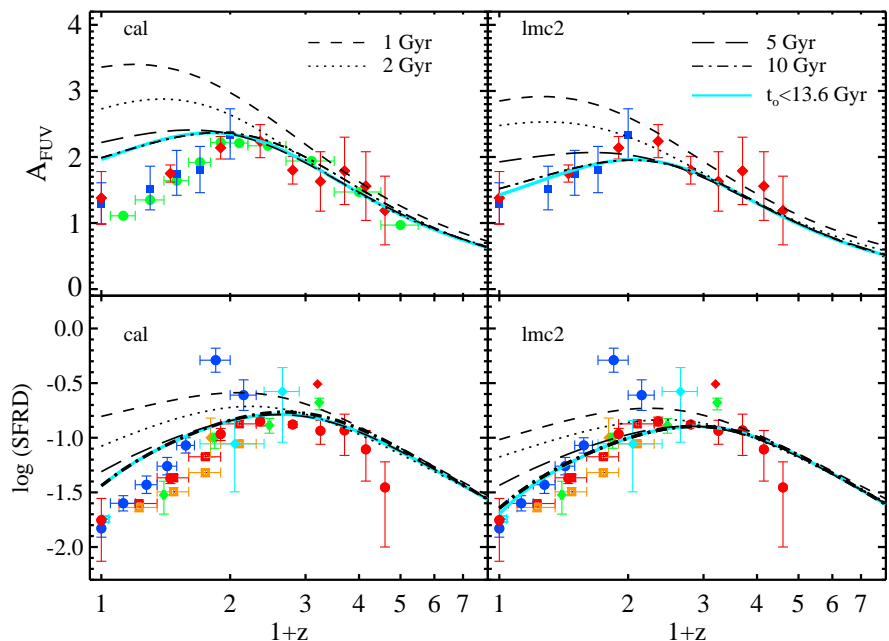}
  \caption{ The $A_{\rm FUV}$($z$) (\emph{top panel}) and SFRD($z$) ( in units of M$_{\odot}$ yr$^{-1}$ Mpc$^{-3}$; \emph{bottom panel}) 
for different ages of the stellar populations contributing in the convolution integral (Eq.\ref{Eq.convolution}) for our `cal' and `lmc2' models. 
The time $t_0<13.6$Gyr is the age with our fiducial $z_{max}=\infty$ limit. Different data points plotted in the \emph{top} and 
\emph{bottom panel} are same as in the Fig.~\ref{fig.dust_shade_all} and  Fig.~\ref{fig.sfrd_shade_all}, respectively.
Legends are scattered over entire plot.}
\label{fig.tmax}
\end{figure*}
%

\subsection{Maximum redshift $z_{max}$ in the convolution}
The convolution integral (Eq.~\ref{Eq.convolution}) gives the $\rho_{\nu}(z_0)$ by 
convolving the SFRD($z$) with the intensity output of instantaneous burst of star formation 
which has occurred at epoch $t$ corresponding to redshift $z$. The population synthesis 
models give the specific intensity $l_{\nu}(\tau)$ where $\tau$ is an age of the stellar 
population which goes through the starburst at $z$ and shining at $z_0$. 
In the calculations discussed above we have taken the maximum redshift $z_{max}=\infty$, 
which means, in principle, the $\rho_{\nu}(z_0)$ will have the contribution from very old 
stars up to the ages of, $t_{max}$, equal to the light travel time from the Big-Bang to 
redshift $z=z_0$ which is less than 13.6 Gyr depending on $z_0$. Note that the actual 
contribution from very old stars (age $>$10 Gyr) which went through the starburst at 
time $t$ is negligible because the SFRD($z$) at $z$ corresponding to these earlier 
epochs $t$ is negligible.
Here, we are investigating the validity of $z_{max}=\infty$ assumption and the effect 
of taking different values of $z_{max}$ (or corresponding maximum age of galaxy $t_{max}$) 
on our derived quantities like $A_{\rm FUV}$ and SFRD$(z)$ mainly at low redshifts 
where the age of the universe is large. 
If we take sufficiently small values of $t_{max}$, a contribution from the old stellar 
population, which shines at the optical and NIR wavelengths, will be less. Therefore, 
galaxies will be bluer than one expects. It requires a large dust attenuation to make 
them red and match the $\rho_{\nu}$ measurements at the NIR wavelengths. With this small 
$t_{max}$, if we follow the progressive fitting method described in 
section~\ref{sec.pfmethod}, we get large $A_{\rm FUV}$($z$) and hence large SFRD($z$) at low 
redshifts. This is demonstrated in the case of `cal' and `lmc2' model in 
Fig.~\ref{fig.tmax} where we show the SFRD($z$) and $A_{\rm FUV}(z)$ determined by 
stopping the convolution integral after a time $t_{max}=t_0$ for different values of 
$t_0$ ranging from 1 Gyr to 10 Gyr. We take $t_{max}$ equal to the age of the universe 
when the age is smaller than $t_0$. We see similar trends for all five models but for the 
sake of presentation we show results in Fig.~\ref{fig.tmax} only for `cal' and `lmc2' model.
It is clear from the Fig.~\ref{fig.tmax} that lower the value of $t_{max}$, higher will 
be the values of inferred SFRD($z$) and $A_{\rm FUV}$($z$). 

In Fig.~\ref{fig.tmax}, we also plot the independent measurements shown in 
Fig.~\ref{fig.dust_shade_all} and  Fig.~\ref{fig.sfrd_shade_all}. These measurements of the 
$A_{\rm FUV}$ shows that it increases from $z=0$ to a peak at $z\sim 1$ and then decreases at 
higher $z$ \citep{Burgarella13, Cucciati12, Takeuchi05}. To get such a shape of $A_{\rm FUV}$, 
as shown in Fig.~\ref{fig.tmax},one needs $t_{max}>5$ Gyr. We find that the 
SFRD($z$) and $A_{\rm FUV}$($z$) are insensitive to $t_{max}$ when it is $\ge 10$ Gyr. 
Therefore, to get the SFRD($z$) and $A_{\rm FUV}$($z$) consistent with the trend seen in 
different independent observations one needs to consider the stellar population ages 
$\ge$ 10 Gyr which is consistent with $z_{max}=\infty$ and the estimated age 11 Gyr 
of Milky-way \citep{Krauss03}.
\subsection{Metallicity}
\begin{figure*}
\centering
  \includegraphics[bb=65 370 545 710,width=12.6cm,keepaspectratio,clip=true]{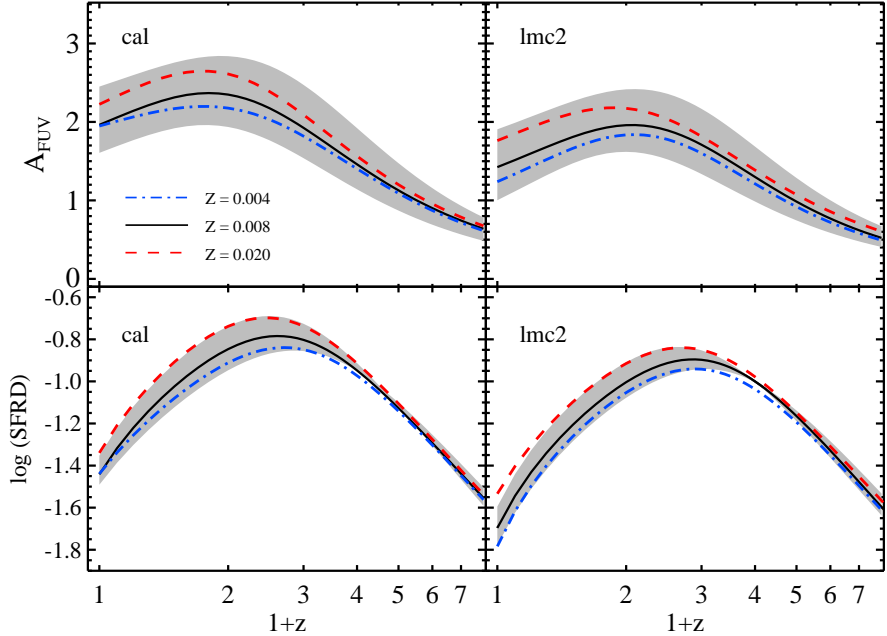}
  \caption{The $A_{\rm FUV}$($z$) (\emph{top panel}) and SFRD($z$) 
(in units of M$_{\odot}$ yr$^{-1}$ Mpc$^{-3}$; \emph{bottom panel}) 
for different metallicity  $Z=0.004$, $Z=0.008$ 
(fiducial) and $Z=0.020$ of the stellar populations for our `cal' and `lmc2' models. 
Gray shaded region represent the range covered in the $A_{\rm FUV}$ 
and SFRD determined using the high and low models for our fiducial $Z=0.008$
(same as shown in Fig.~\ref{fig.dust_shade_all} and Fig.~\ref{fig.sfrd_shade_all}). }
\label{fig.metal}
\end{figure*}

Another source of possible uncertainty in determining the SFRD($z$) and $A_{\rm FUV}$($z$) 
can be related to the redshift evolution of metallicity, which we do not consider. We use 
constant $Z=0.008$ metallicity for all redshifts. HM12 and \citet{Madau14} use the 
metallicity evolution as $Z(z)=Z_{\odot}10^{-0.15z}$ suggested by \citet{Kewley07} where 
$Z_{\odot}=0.020$. This evolution gives $Z=0.008$ at $z=2.6$. To see the effect of using 
different metallicity, we determine the SFRD($z$) and $A_{\rm FUV}$($z$) for metallicity 
$Z=0.020$ and $Z=0.004$. Our results are plotted in Fig.~\ref{fig.metal} for the `cal' and 
`lmc2' models. We also show the range covered by the $A_{\rm FUV}$ and SFRD when obtained for 
respective high and low models for our fiducial metallicity $Z=0.008$ 
(same as in the Fig.~\ref{fig.dust_shade_all} and  Fig.~\ref{fig.sfrd_shade_all}). Note that 
the low and high models use 1-$\sigma$ low and high fit through $\rho_{\rm FUV}$ 
measurements used to determine the $A_{\rm FUV}$ and SFRD. It is clear from 
Fig.~\ref{fig.metal} that the higher (lower) metallicity gives higher (lower) 
$A_{\rm FUV}$($z$) and SFRD($z$). We see similar trends for all five models but for the sake 
of presentation we show only for `cal' and `lmc2' model. The difference between the 
$A_{\rm FUV}$($z$) obtained for these metallicities (as high as factor 5 in $Z$), are well 
within the allowed range of $A_{\rm FUV}$ and SFRD obtained using our fiducial $Z=0.008$. 
This suggests that the scatter in the $A_{\rm FUV}$($z$) and SFRD($z$) arising due to change 
in metallicity (as high as factor 5 ) is smaller than the scatter one gets in the 
$A_{\rm FUV}$($z$) and SFRD($z$) due to scatter in $\rho_{\rm FUV}$ measurements at a 
constant metallicity. 

The analysis presented here shows that the effect of the metallicity evolution is sub-dominant 
as compared to those arising from uncertainties in $\rho_{\rm FUV}$ measurements. Therefore, 
our assumption of constant metallicity is valid and compatible with the current $\rho_{\nu}$ 
measurements. 
%
%

\subsection{IMF and $L_{min}$}
In this paper, we consider the \citet{Salpeter55} IMF in the population synthesis model with 
stellar mass range from 0.1 to 100$M_{\odot}$. Even though there are other IMFs like 
\citet{Kroupa03}, \citet{Chabrier03} and  \citet{Baldry03}, since most of the star formation 
rates reported in the literature use Salpeter IMF, we also prefer it for our work. Note that, 
the different IMF can change the combination of $A_{\rm FUV}$($z$) and SFRD($z$) but the 
fact that the progressive fitting method we use to get these combinations will ensure that 
the emissivities and EBL will remain the same in UV to NIR wavelengths (the wavelengths where 
we have multi-wavelength multi-epoch luminosity functions). However, the different IMF and 
hence different $A_{\rm FUV}$($z$) and SFRD($z$) will give different FIR emissivity and 
FIR part of the EBL \citep[see for e.g][]{Primack01}.

The luminosity densities calculated from the observed luminosity function depend on the 
values of $L_{min}$ and the faint end slope $\alpha$. For most of the $\rho_{\nu}$ used here, 
we use $L_{min}=0$. In Table~\ref{lmin_table} of Appendix, we list $\alpha$ and $L_{min}$ 
values used to get $\rho_{\nu}$. We also list the percentage decrease in the $\rho_{\nu}$ if 
we use $L_{min}=0.01L^*$ and $L_{min}=0.03L^*$ instead of the fiducial $L_{min}$ (see the 
column labeled as $\Delta_1$ and $\Delta_2$ in Table~\ref{lmin_table}). The difference is 
large for the small $\alpha$ (i.e, $\alpha<-1.3$). The value $L_{min}=0.01L^*$ is used by 
HM12 in their UV background calculations. The value of $L_{min}=0.03L^*$ is used by 
\citet{Madau14} to get the SFRD. It can be directly seen from the Table~\ref{lmin_table} that 
the choice of $L_{min}$ between 0 to 0.03$L^*$ can change the $\rho_{\nu}$ upto 30\%. This 
difference is large for the $\rho_{\nu}$ measurements of \citet{Tresse07} at high $z$ and 
higher wavebands where $\alpha$ is small. Because of the sensitivity limit of  
survey, \citet{Tresse07} could not determine $\alpha$ for $z>1.2$. Therefore, at high $z$, 
the $\alpha$ is extrapolated from the low redshift measurements. However, note that the 
errors estimated on the $\rho_{\nu}$ values by \citet{Tresse07} include uncertainties 
arising from the differer values of $\alpha$ which is larger than the difference due to 
$L_{min}$ values mentioned here.

By increasing the values of $L_{min}$, we can get the lower $\rho_{\nu}$ which will give rise 
to the EBL with lower intensity. Since our EBL generated using the $\rho_{\nu}$ with 
$L_{min}=0$ passes through the lower limits on the local EBL obtained from the IGL 
measurements in the FUV to NIR bands (see Fig.~\ref{fig.ebl}), we do not consider the higher 
values of $L_{min}$. 
 
%
\section{Summary}\label{sec.sum}
In this paper, we estimate the extragalactic background light (EBL) which is consistent with 
the observed multi-wavelength and multi-epoch luminosity functions upto $z\sim 8$.
To achieve that we introduce a novel method which determines the unique combination of the 
dust attenuation magnitude at FUV band, $A_{\rm FUV}(z)$, and the star formation rate density, 
SFRD$(z)$, for a given extinction curve. It allows us to investigate the mean extinction curve 
which can be used to determine the global average quantities like EBL, SFRD$(z)$ and 
$A_{\rm FUV}(z)$. The main results of our work are summarized below.

\begin{enumerate}
\item We introduce a `progressive fitting method' which uses multi-wavelength and multi-epoch 
luminosity functions to determine a unique combination $A_{\rm FUV}(z)$ and SFRD$(z)$ for a 
given extinction curve. The combination of $A_{\rm FUV}(z)$ and  SFRD$(z)$, by construct, 
reproduces the emissivity consistent with the observed luminosity functions. 
\item We compiled the observed luminosity functions from the FUV to K band and upto 
$z \sim 8$. Using this we determine the combinations of $A_{\rm FUV}(z)$ and SFRD$(z)$ from 
the `progressive fitting method' for a set of well known extinction curves observed for 
Milky-Way, Small Megallenic Clouds (SMC), Large Megallenic Clouds (LMC), LMC supershell 
(LMC2) and for the nearby starburst galaxies given by \citet{Calzetti}. 
\item With the help of these combinations of $A_{\rm FUV}(z)$ and SFRD$(z)$, for each 
extinction curve, we calculate the average energy absorbed by the inter-stellar dust in the 
UV to NIR wavelengths. This allowed us to estimate the NIR to FIR emissivity using the 
principle of energy conservation and the IR emission templates of local galaxies.
\item Out of all five extinction curves, we find that the $A_{\rm FUV}(z)$, SFRD$(z)$ and 
local emissivity obtained using LMC2 extinction curve reproduces different independent 
measurements. This enables us to conclude that, out of five well measured extinction curves 
for nearby galaxies, the average extinction curve which is applicable for galaxies over wide 
range of redshifts is most likely to be similar to LMC2 extinction curve. 
\item We use the emissivity obtained for each extinction curve from UV to IR wavelengths and 
calculate the EBL for each. We compare these with the different EBL estimates reported in 
the literature and with the lower and upper limits placed on the local EBL from various 
observations. 
\item For different EBLs estimated here, we calculate the optical depths, $\tau_{\gamma}$, 
encountered by the high energy $\gamma$-rays due to electron positron pair production upon 
collision with the EBL photons. We compare the $\tau_{\gamma}$ computed for our EBL with 
those from different EBL estimates reported in the literature and with the measurements of 
\citet{GammaSci}. We also calculate the $\gamma$-ray horizon and compare with recently 
reported measurements of \citet{Dominguez13}.
\item We find that the IR part of the local EBL and the corresponding $\gamma$-ray horizon in 
TeV energies estimated using the SMC extinction curve are inconsistent with various 
measurements. However, these measurements are consistent with results obtained from all other 
extinction curves. 
\item We discuss the uncertainties in $A_{\rm FUV}(z)$, SFRD$(z)$ and the EBL estimates 
related to the standard assumptions like metallicity, faint end slope of the luminosity 
function and age of the stellar population contributing to the emissivity.
\end{enumerate}
 
We fit the $A_{\rm FUV}(z)$ and SFRD($z$) using a functional form given in 
Eq.~\ref{Eq.fit_form} and the fitting parameters obtained for each extinction curve are 
provided in Table~\ref{dust_parm} \& Table~\ref{sfrd_parm}. From the very good agreement with 
various measurements we conclude that the LMC2 extinction curve should be used to
calculate the global averaged quantities like the EBL, SFRD and $A_{\rm FUV}$. 

The `progressive fitting method' used here to get the $A_{\rm FUV}(z)$ and SFRD$(z)$ requires 
luminosity functions observed over different wavebands and redshifts. Therefore, the surveys 
like \citet{Tresse07} and \citet{Ilbert05} are very important which determine the luminosity 
functions uniformly over large redshifts and different wavebands. Currently our method is 
limited by the lack of good observations in different wavebands and at high redshifts.

\section*{Acknowledgments} 
We wish to thank Lawrence Tresse, Kari Helgason, Peter Behroozi, Marco Ajello, Hamsa Padmanabhan,
 Tirthankar Roy Choudhury and Sunder Sahayanathan for
providing relevant data and useful discussions. We thank anonymous referee for useful comments
which helped us to improve the paper.
VK thanks CSIR for providing support for this work. 

%
\def\apj{ApJ}%
\def\mnras{MNRAS}%
\def\aap{A\&A}%
\def\apjl{ApJ}
\def\aj{AJ}
\def\physrep{PhR}
\def\apjs{ApJS}
\def\pasa{PASA}
\def\pasj{PASJ}
\def\pasp{PASP}
\def\nat{Natur}
\def\araa{AR\&A}
\bibliographystyle{apj}
\bibliography{vikrambib}

\appendix
\section{Additional Figure and Table }\label{appendix}
\begin{figure}
\centering
  \includegraphics[bb=30 20 590 770, width=15cm, keepaspectratio,clip=true]{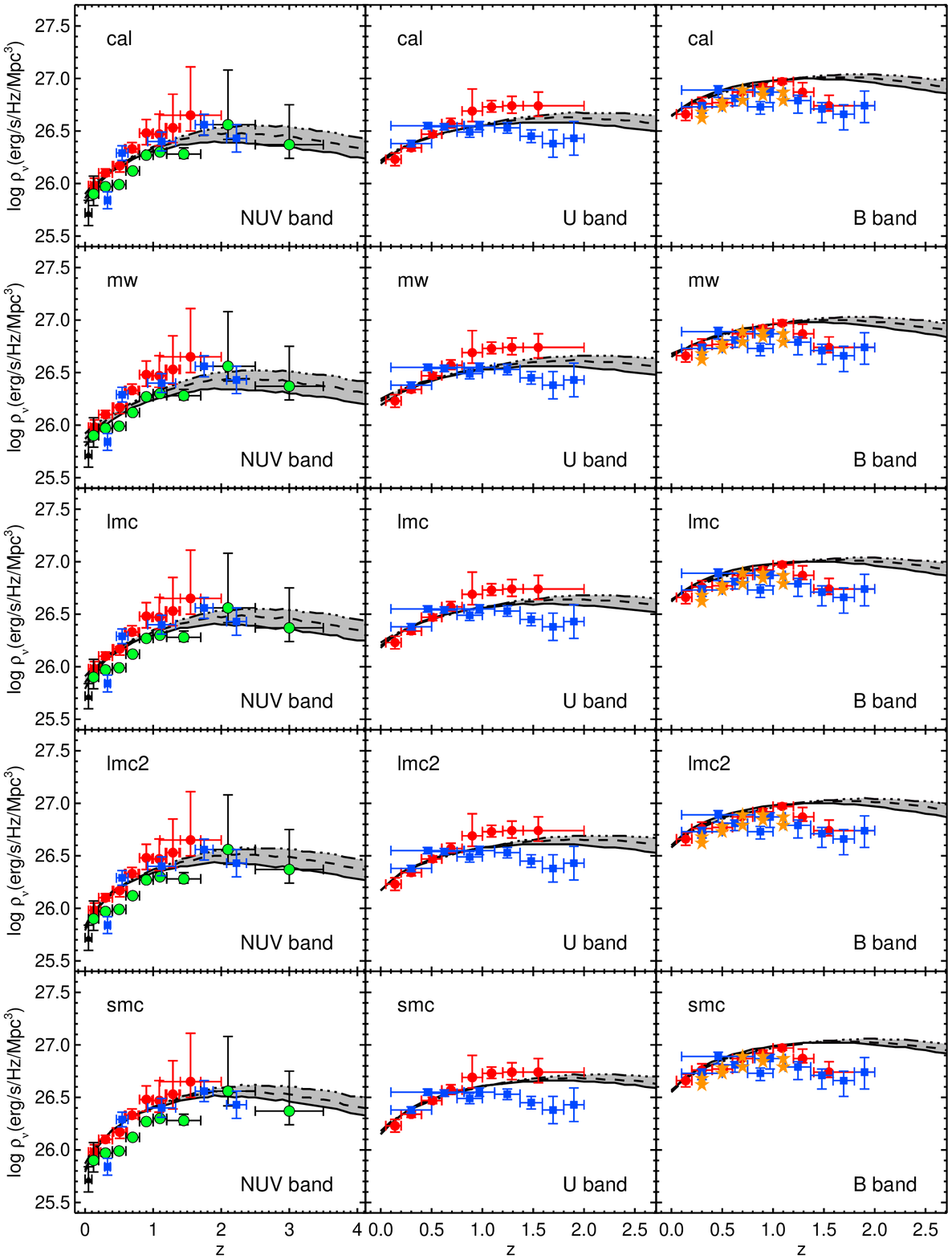}
  \caption{\footnotesize {NUV, U and V band comoving luminosity density calculated using best fit combination of SFRD($z$) and $A_{\rm FUV}(z)$
obtained using different extinction curves $k(\nu)$. For the references and plotting symbols see Table~\ref{lf_data}.
Solid, dashed and dot-dashed lines represent values of the best fit $\rho_{\nu}$ calculated for the SFRD($z$) obtained using the low, 
median and high models, respectively. From \emph{top} to \emph{bottom} extinction curve used are \citet{Calzetti}, Milky-Way, 
 LMC, LMC2 and SMC, respectively.}}
\label{fig.A1}
\end{figure}
\begin{figure*}
\centering
\setcounter{figure}{16}
   \includegraphics[bb=30 20 590 770, width=15cm, keepaspectratio,clip=true]{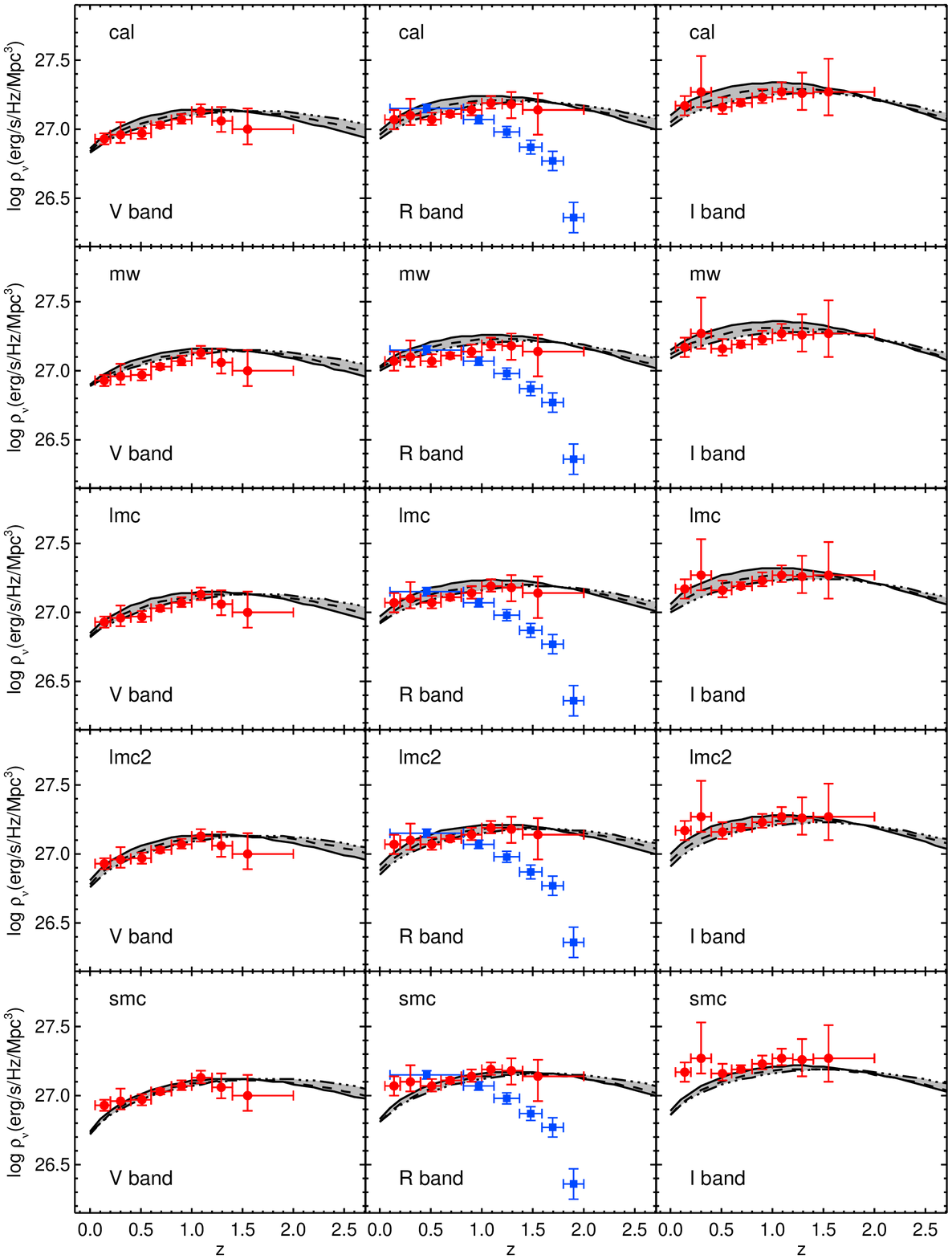}
  \caption{  \small{... continued (for V, R and I band). For the references and plotting symbols see Table~\ref{lf_data}.
Solid, dashed and dot-dashed lines represent values of the best fit $\rho_{\nu}$ calculated for the SFRD($z$) obtained using the low, 
median and high models, respectively. From \emph{top} to \emph{bottom} extinction curve used are \citet{Calzetti}, Milky-Way, 
 LMC, LMC2 and SMC, respectively.}} 
\end{figure*}
\begin{figure*}
\centering
\setcounter{figure}{16}
   \includegraphics[bb=30 20 590 770, width=15cm, keepaspectratio,clip=true]{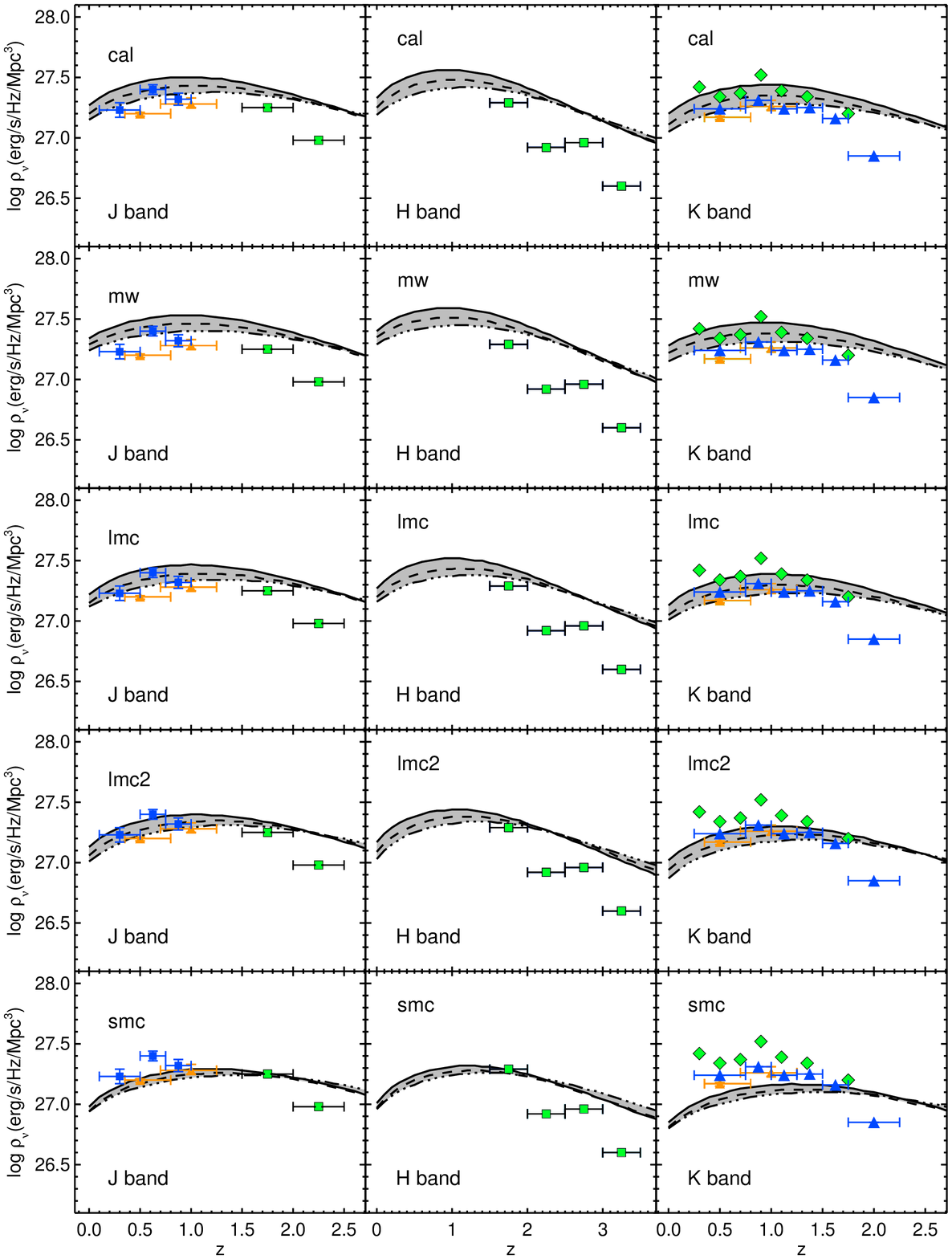}
  \caption{ \small{... continued (for J, H and K band). For the references and plotting symbols see Table~\ref{lf_data}.
Solid, dashed and dot-dashed lines represent values of the best fit $\rho_{\nu}$ calculated for the SFRD($z$) obtained using the low, 
median and high models, respectively. From \emph{top} to \emph{bottom} extinction curve used are \citet{Calzetti}, Milky-Way, 
 LMC, LMC2 and SMC, respectively.}}
\end{figure*}
%
%
\begin{table*}
\caption{Observations of the Galaxy Luminosity Function}
\label{lmin_table} 
\centering
\setcounter{table}{4}
\begin{tabular}{l l l l l l l | l l l l l l l}
\hline\hline
 Reference & band & z & $\alpha$ & L$_{min}$ & $\Delta_1$ & $\Delta_2$ & Reference & band & z &  $\alpha$ & L$_{min}$ & $\Delta_1$ & $\Delta_2$   \\
           &      &   &          & (L$^*$)& \%          & \%        &           &      &   &          & (L$^*$)& \%          & \%           \\
\hline
\hline
\citet{Cucciati12}    & FUV & 0.125 & -1.05 & 0    &  1 &  3 & \citet{Tresse07}      &  U  & 0.14  & -1.05 & 0    &  1 &  3 \\
\citet{Tresse07}      & FUV & 0.14  & -1.13 & 0    &  1 &  4 & \citet{Dahlen05}      &  U  & 0.3   & -1.31 & 0    &  4 &  9 \\
\citet{Cucciati12}    & FUV & 0.3   & -1.17 & 0    &  2 &  5 & \citet{Tresse07}      &  U  & 0.3   & -1.17 & 0    &  2 &  5 \\
\citet{Schiminovich05}& FUV & 0.3   & -1.19 & 0    &  2 &  6 & \citet{Dahlen05}      &  U  & 0.46  & -1.2  & 0    &  2 &  6 \\
\citet{Tresse07}      & FUV & 0.3   & -1.6  & 0    & 15 & 25 & \citet{Tresse07}      &  U  & 0.51  & -1.17 & 0    &  2 &  5 \\
\citet{Cucciati12}    & FUV & 0.5   & -1.07 & 0    &  1 &  3 & \citet{Dahlen05}      &  U  & 0.625 & -1.31 & 0    &  4 &  9 \\
\cite{Schiminovich05} & FUV & 0.5   & -1.55 & 0    & 12 & 22 & \citet{Tresse07}      &  U  & 0.69  & -1.27 & 0    &  3 &  8 \\
\citet{Tresse07}      & FUV & 0.51  & -1.6  & 0    & 15 & 25 & \citet{Dahlen05}      &  U  & 0.875 & -1.31 & 0    &  4 &  9 \\
\citet{Tresse07}      & FUV & 0.69  & -1.6  & 0    & 15 & 25 & \citet{Tresse07}      &  U  & 0.9   & -1.44 & 0    &  8 & 15 \\
\citet{Cucciati12}    & FUV & 0.7   & -0.9  & 0    &  0 &  1 & \citet{Dahlen05}      &  U  & 0.97  & -1.2  & 0    &  2 &  6 \\
\citet{Schiminovich05}& FUV & 0.7   & -1.6  & 0    & 15 & 25 & \citet{Tresse07}      &  U  & 1.09  & -1.51 & 0    & 10 & 19 \\
\citet{Cucciati12}    & FUV & 0.9   & -0.85 & 0    &  0 &  1 & \citet{Dahlen07}      &  U  & 1.245 & -1.2  & 0    &  2 &  6 \\
\citet{Tresse07}      & FUV & 0.9   & -1.6  & 0    & 15 & 25 & \citet{Tresse07}      &  U  & 1.29  & -1.59 & 0    & 15 & 25 \\
\citet{Schiminovich05}& FUV & 1     & -1.6  & 0    & 15 & 25 & \citet{Dahlen05}      &  U  & 1.48  & -1.2  & 0    &  2 &  6 \\
\citet{Tresse07}      & FUV & 1.09  & -1.6  & 0    & 15 & 25 & \citet{Tresse07}      &  U  & 1.55  & -1.68 & 0    & 21 & 32 \\
\citet{Cucciati12}    & FUV & 1.1   & -0.91 & 0    &  0 &  2 & \citet{Dahlen05}      &  U  & 1.695 & -1.2  & 0    &  2 &  6  \\
\citet{Dahlen07}      & FUV & 1.125 & -1.48 & 0    &  9 & 17 & \citet{Dahlen05}      &  U  & 1.9   & -1.2  & 0    &  2 &  6 \\
\citet{Tresse07}      & FUV & 1.29  & -1.6  & 0    & 15 & 25 & \citet{Tresse07}      &  B  & 0.14  & -1.09 & 0    &  1 &  4 \\
\citet{Cucciati12}    & FUV & 1.45  & -1.09 & 0    &  1 &  4 & \cite{Dahlen05}       &  B  & 0.3   & -1.37 & 0    &  5 & 11 \\
\citet{Tresse07}      & FUV & 1.55  & -1.6  & 0    & 15 & 25 & \citet{Tresse07}      &  B  & 0.3   & -1.15 & 0    &  2 &  5 \\
\citet{Dahlen07}      & FUV & 1.75  & -1.48 & 0    &  9 & 17 & \citet{Dahlen05}      &  B  & 0.46  & -1.28 & 0    &  3 &  8 \\
\cite{Schiminovich05} & FUV & 2.0   & -1.49 & 0    & 10 & 18 & \citet{Tresse07}      &  B  & 0.51  & -1.22 & 0    &  2 &  6 \\
\citet{Cucciati12}    & FUV & 2.1   & -1.3  & 0    &  4 &  9 & \citet{Dahlen05}      &  B  & 0.625 & -1.37 & 0    &  5 & 11 \\
\citet{Dahlen07}      & FUV & 2.22  & -1.48 & 0    &  9 & 17 & \citet{Tresse07}      &  B  & 0.69  & -1.12 & 0    &  1 &  4 \\
\citet{Reddy09}$^a$   & FUV & 2.3   & -1.73 & 0.01 &  0 & 15 & \citet{Dahlen05}      &  B  & 0.875 & -1.37 & 0    &  5 & 11 \\
\citet{Schiminovich05}& FUV & 2.9   & -1.47 & 0    &  9 & 16 & \citet{Tresse07}      &  B  & 0.9   & -1.33 & 0    &  4 & 10 \\
\citet{Cucciati12}    & FUV & 3     & -1.5  & 0    & 10 & 18 & \citet{Dahlen05}      &  B  & 0.97  & -1.28 & 0    &  3 &  8 \\
\citet{Reddy09}$^a$   & FUV & 3.05  & -1.73 & 0.01 &  0 & 15 & \citet{Tresse07}      &  B  & 1.09  & -1.40 & 0    &  6 & 13 \\
\citet{Bouwens11}$^b$ & FUV & 3.8   & -1.73 & 0.01 &  0 & 15 & \citet{Dahlen05}      &  B  & 1.245 & -1.28 & 0    &  3 &  8  \\
\citet{Cucciati12}    & FUV & 4     & -1.5  & 0    & 10 & 18 & \citet{Tresse07}      &  B  & 1.29  & -1.48 & 0    &  9 & 17 \\
\citet{Bouwens11}$^b$ & FUV & 5     & -1.66 & 0.01 &  0 & 13 & \citet{Dahlen05}      &  B  & 1.48  & -1.28 & 0    &  3 &  8 \\
\citet{Bouwens11}$^c$ & FUV & 5.9   & -1.74 & 0.01 &  0 & 16 & \citet{Tresse07}      &  B  & 1.55  & -1.57 & 0    & 14 & 23 \\
\citet{Bouwens11}$^b$ & FUV & 6.8   & -2.01 & 0.01 &  0 & 27 & \citet{Dahlen05}      &  B  & 1.695 & -1.28 & 0    &  3 &  8  \\
\citet{Bouwens11}$^d$ & FUV & 8     & -1.91 & 0.01 &  0 & 22 & \citet{Dahlen05}      &  B  & 1.9   & -1.28 & 0    &  3 &  8 \\
\citet{Wyder05}       & NUV & 0.055 & -1.16 & 0    &  2 &  5 & \citet{Faber07}       &  B  & 0.3   & -1.3  & 0.01 &  0 &  5 \\
\citet{Cucciati12}    & NUV & 0.125 & -1.08 & 0    &  1 &  4 & \citet{Faber07}       &  B  & 0.5   & -1.3  & 0.01 &  0 &  5 \\
\citet{Tresse07}      & NUV & 0.14  & -1.32 & 0    &  4 &  9 & \citet{Faber07}       &  B  & 0.7   & -1.3  & 0.01 &  0 &  5 \\
\citet{Cucciati12}    & NUV & 0.3   & -1.02 & 0    &  1 &  3 & \citet{Faber07}       &  B  & 0.9   & -1.3  & 0.01 &  0 &  5 \\
\citet{Tresse07}      & NUV & 0.3   & -1.32 & 0    &  4 &  9 & \citet{Faber07}       &  B  & 1.1   & -1.3  & 0.01 &  0 &  5 \\
\citet{Dahlen07}      & NUV & 0.33  & -1.39 & 0    &  6 & 12 & \citet{Tresse07}      &  V  & 0.14  & -1.15 & 0    &  2 &  5 \\
\citet{Cucciati12}    & NUV & 0.5   & -1.08 & 0    &  1 &  4 & \citet{Tresse07}      &  V  & 0.3   & -1.21 & 0    &  2 &  6 \\
\citet{Tresse07}      & NUV & 0.51  & -1.32 & 0    &  4 &  9 & \citet{Tresse07}      &  V  & 0.51  & -1.35 & 0    &  5 & 11 \\
\citet{Dahlen07}      & NUV & 0.545 & -1.39 & 0    &  6 & 12 & \citet{Tresse07}      &  V  & 0.69  & -1.35 & 0    &  5 & 11 \\
\citet{Tresse07}      & NUV & 0.69  & -1.32 & 0    &  4 &  9 & \citet{Tresse07}      &  V  & 0.9   & -1.50 & 0    & 10 & 18 \\
\citet{Cucciati12}    & NUV & 0.7   & -0.95 & 0    &  0 &  2 & \citet{Tresse07}      &  V  & 1.09  & -1.57 & 0    & 14 & 23 \\
\citet{Cucciati12}    & NUV & 0.9   & -0.81 & 0    &  0 &  1 & \citet{Tresse07}      &  V  & 1.29  & -1.65 & 0    & 19 & 30 \\
\citet{Tresse07}      & NUV & 0.9   & -1.32 & 0    &  4 &  9 & \citet{Tresse07}      &  V  & 1.55  & -1.74 & 0    & 27 & 38 \\
\citet{Tresse07}      & NUV & 1.09  & -1.32 & 0    &  4 &  9 & \citet{Tresse07}      &  R  & 0.14  & -1.16 & 0    &  2 &  5 \\
\citet{Cucciati12}    & NUV & 1.1   & -0.88 & 0    &  0 &  1 & \citet{Tresse07}      &  R  & 0.3   & -1.27 & 0    &  3 &  8 \\
\citet{Dahlen07}      & NUV & 1.125 & -1.39 & 0    &  6 & 12 & \citet{Dahlen05}      &  R  & 0.46  & -1.30 & 0    &  4 &  9 \\
\citet{Tresse07}      & NUV & 1.29  & -1.32 & 0    &  4 &  9 & \citet{Tresse07}      &  R  & 0.51  & -1.42 & 0    &  7 & 14 \\
\citet{Cucciati12}    & NUV & 1.45  & -1.05 & 0    &  1 &  3 & \citet{Tresse07}      &  R  & 0.69  & -1.41 & 0    &  7 & 13 \\
\citet{Tresse07}      & NUV & 1.55  & -1.32 & 0    &  4 &  9 & \citet{Tresse07}      &  R  & 0.9   & -1.53 & 0    & 11 & 20 \\
\citet{Dahlen07}      & NUV & 1.75  & -1.39 & 0    &  6 & 12 & \citet{Dahlen05}      &  R  & 0.97  & -1.30 & 0    &  4 &  9 \\
\citet{Cucciati12}    & NUV & 2.1   & -1.16 & 0    &  2 &  5 & \citet{Tresse07}      &  R  & 1.09  & -1.60 & 0    & 15 & 25 \\
\citet{Dahlen07}      & NUV & 2.225 & -1.39 & 0    &  6 & 12 & \citet{Dahlen05}      &  R  & 1.245 & -1.30 & 0    &  4 &  9 \\
\citet{Cucciati12}    & NUV & 3     & -1.15 & 0    &  2 &  5 & \citet{Tresse07}      &  R  & 1.29  & -1.68 & 0    & 21 & 32 \\
\hline
\hline
\end{tabular}
\hfill
\label{lmin_data}
\begin{flushleft}
\footnotesize {Wavelengths at different bands are: FUV=0.15$\mu$m, NUV=0.28$\mu$m, U=0.365$\mu$m, B=0.445$\mu$m, 
V=0.551$\mu$m, R=0.658$\mu$m\\} 
\footnotesize {$\Delta_1=(a_0-a_1)/a_0$  and $\Delta_2=(a_0-a_2)/a_0$ where $a_0=\Gamma(\alpha+2, L_{min}/L^*)$, 
$a_1=\Gamma(\alpha+2, 0.01)$, and $a_2=\Gamma(\alpha+2, 0.03)$ } \\
\footnotesize {$^a$ The corresponding wavelength is 1700\AA.}\\
\footnotesize {$^b$ The corresponding wavelength is 1600\AA.}
\footnotesize {$^c$ The corresponding wavelength is 1350\AA.}
\footnotesize {$^d$ The corresponding wavelength is 1750\AA.}\\

\end{flushleft}
\end{table*}
\begin{table*}
\caption{...(continue) Observations of the Galaxy Luminosity Function}
\label{lmin_table} 
\centering
\begin{tabular}{l l l l l l l| l l l l l l l}
\hline\hline
 Reference & band & z & $\alpha$ & L$_{min}$ & $\Delta_1$ & $\Delta_2$ & Reference & band & z &  $\alpha$ & L$_{min}$ & $\Delta_1$ & $\Delta_2$ \\
           &      &   &          & (L$^*$)& \%          & \%        &           &      &   &          & (L$^*$)& \%          & \%         \\
\hline
\hline
\citet{Dahlen05}      &  R  & 1.48  & -1.30 & 0     &  4 &  9 & SM13                   &  H  & 1.75  &-1.30  & 0.01 &  0 & 5    \\
\citet{Tresse07}      &  R  & 1.55  & -1.77 & 0     & 29 & 42 & SM13                   &  H  & 2.25  &-1.23  & 0.01 &  0 & 4    \\
\citet{Dahlen05}      &  R  & 1.695 & -1.30 & 0     &  4 &  9 & SM13                   &  H  & 2.75  &-1.11  & 0.01 &  0 & 2   \\ 
\citet{Dahlen05}      &  R  & 1.9   & -1.30 & 0     &  4 &  9 & SM13                   &  H  & 3.35  &-1.30  & 0.01 &  0 & 5   \\
\citet{Tresse07}      &  I  & 0.14  & -1.19 & 0     &  2 &  6 & \citet{Smith09}        &  K  & 0.155 & -0.81 & 0.01 &  0 & 1   \\
\citet{Tresse07}      &  I  & 0.3   & -1.32 & 0     &  4 &  9 & \citet{Arnouts07}      &  K  & 0.30  & -1.1  & 0.01 &  0 & 2   \\
\citet{Tresse07}      &  I  & 0.51  & -1.47 & 0     &  9 & 16 & \citet{Arnouts07}      &  K  & 0.50  & -1.1  & 0.01 &  0 & 2   \\
\citet{Tresse07}      &  I  & 0.69  & -1.41 & 0     &  7 & 13 & \citet{Cirasuolo07}    &  K  & 0.5   & -0.99 & 0.01 &  0 & 1   \\
\citet{Tresse07}      &  I  & 0.9   & -1.52 & 0     & 11 & 19 & \citet{Pozzetti03}     &  K  & 0.5   & -1.25 & 0.02 & -2 & 2   \\
\citet{Tresse07}      &  I  & 1.09  & -1.59 & 0     & 15 & 25 & \citet{Arnouts07}      &  K  & 0.70  & -1.1  & 0.01 &  0 & 2   \\
\citet{Tresse07}      &  I  & 1.29  & -1.67 & 0     & 20 & 31 & \citet{Cirasuolo07}    &  K  & 0.875 & -1.00 & 0.01 &  0 & 1   \\
\citet{Tresse07}      &  I  & 1.55  & -1.76 & 0     & 28 & 41 & \citet{Arnouts07}      &  K  & 0.90  & -1.1  & 0.01 &  0 & 2   \\
\citet{Dahlen05}      &  J  & 0.3   & -1.48 & 0     &  9 & 17 & \citet{Pozzetti03}     &  K  & 1     & -0.98 & 0.02 &  0 & 0   \\
\citet{Pozzetti03}    &  J  & 0.5   & -1.22 & 0.02  & -2 &  1 & \citet{Cirasuolo07}    &  K  & 0.875 & -1.00 & 0.01 &  0 & 1   \\
\citet{Dahlen05}      &  J  & 0.625 & -1.48 & 0     &  9 & 17 & \citet{Arnouts07}      &  K  & 1.10  & -1.1  & 0.01 &  0 & 2   \\
\citet{Dahlen05}      &  J  & 0.875 & -1.48 & 0     &  9 & 17 & \citet{Cirasuolo07}    &  K  & 1.125 & -0.94 & 0.01 &  0 & 1   \\
\citet{Pozzetti03}    &  J  & 1     & -0.86 & 0.02  &  0 &  0 & \citet{Arnouts07}      &  K  & 1.35  & -1.1  & 0.01 &  0 & 2   \\
SM13 $^d$             &  J  & 1.75  & -1.24 & 0.01  &  0 &  4 & \citet{Cirasuolo07}    &  K  & 1.375 & -0.92 & 0.01 &  0 & 1   \\
SM13                  &  J  & 2.25  & -1.12 & 0.01  &  0 &  2 & \citet{Cirasuolo07}    &  K  & 1.625 & -1    & 0.01 &  0 & 1   \\
SM13                  &  J  & 2.75  & -1.17 & 0.01  &  0 &  3 & \citet{Arnouts07}      &  K  & 1.75  & -1.1  & 0.01 &  0 & 2   \\
SM13                  &  J  & 3.35  & -0.92 & 0.01  &  0 &  1 & \cite{Cirasuolo07}     &  K  & 2     & -1    & 0.01 &  0 & 1    \\

\hline                                            
\hline
\end{tabular}
\hfill
\label{lmin_data}
\begin{flushleft}
\footnotesize {Wavelengths at different wavebands are: R=0.658$\mu$m, I=0.806$\mu$m, J=1.27$\mu$m, H=1.63$\mu$m and K=2.2$\mu$m.}\\
\footnotesize {$\Delta_1=(a_0-a_1)/a_0$  and $\Delta_2=(a_0-a_2)/a_0$ where $a_0=\Gamma(\alpha+2, L_{min}/L^*)$, 
$a_1=\Gamma(\alpha+2, 0.01)$, and $a_2=\Gamma(\alpha+2, 0.03)$ } \\
\footnotesize {$^d$ The acronym SM13 corresponds to the reference \citet{Stefanon13} }\\

 \end{flushleft}
\end{table*}
\end{document}